\documentclass[twoside]{article}

\usepackage[sc]{mathpazo} % Use the Palatino font
\usepackage[T1]{fontenc} % Use 8-bit encoding that has 256 glyphs
\usepackage{microtype} % Slightly tweak font spacing for aesthetics

\usepackage[hmarginratio=1:1,top=32mm,columnsep=15pt]{geometry} % Document margins
\usepackage{hyperref} % For hyperlinks in the PDF
\usepackage{graphicx}  

\usepackage[small,labelfont=bf,up,textfont=it,up]{caption} % Custom captions under/above floats in tables or figures

\usepackage{abstract} % Allows abstract customization
\usepackage{titlesec} % Allows customization of titles
\titleformat{\section}[block]{\large\scshape\centering{\Roman{section}.}}{}{1em}{} % Change the look of the section titles 

\usepackage{fancyhdr} % Headers and footers
\title{\vspace{-15mm}\fontsize{18pt}{10pt}\selectfont\textbf{From cosmic ray physics to cosmic ray astronomy: \\Bruno Rossi and the opening \\of new windows on the universe}} % Article title

\author{
\large
\textsc{Luisa Bonolis}\\[2mm] % Your name
\normalsize Via Cavalese 13 -- 00135 Rome, Italy \\ % Your institution
\normalsize \href{mailto:luisa.bonolis@roma1.infn.it}{luisa.bonolis@roma1.infn.it} % Your email address
\vspace{-5mm}
}
\date{}

\begin{document}

\maketitle % Insert title

\thispagestyle{fancy} % All pages have headers and footers

%----------------------------------------------------------------------------------------
%	ABSTRACT
%----------------------------------------------------------------------------------------

\begin{abstract}

\noindent 
Bruno Rossi is considered one of the fathers of modern physics, being also a
pioneer in virtually every aspect of what is today called high-energy
astrophysics.  At the beginning of 1930s he was the pioneer of cosmic ray
research in Italy, and, as one of the leading actors in the study of the
nature and behavior of the cosmic radiation, he witnessed  the birth of particle physics 
and was one of the main investigators in this fields for many years. While
cosmic ray physics moved more and more towards astrophysics, Rossi continued
to be one of the inspirers of this line of research. When outer space became
a reality, he did not hesitate to leap into this new scientific dimension. Rossi's intuition on the
importance of exploiting new technological windows to look at the universe
with new eyes, is a fundamental key to understand the profound unity which
guided his scientific research path up to its culminating moments at the
beginning of 1960s, when his group at MIT performed the first {\it in situ} measurements of the density, speed and direction of the solar wind at the boundary of Earth's magnetosphere, and when he promoted the search for extra-solar sources of X
rays. A visionary idea which eventually led to the breakthrough experiment
which discovered Scorpius X-1 in 1962, and inaugurated X-ray astronomy.

\end{abstract}


%----------------------------------------------------------------------------------------
%	ARTICLE CONTENTS
%----------------------------------------------------------------------------------------

%\begin{multicols}{2} % Two-column layout throughout the main article text

%\lettrine[nindent=0em,lines=3]{L} orem ipsum dolor sit amet, consectetur adipiscing elit.

%------------------------------------------------

%\section{Methods}
\section{Introduction}

General works on the history and culture of physics generally neglect cosmic-ray studies as a  field of research in itself, giving specific accounts on their long and complex history mainly referring to its connection with particle physics on the one side or astrophysics on the other.\footnote{This work was presented  at the conference ``100 Years Cosmic Ray Physics -- Anniversary of the V.F. Hess Discovery''; 6--8 August, Bad Saarow/Pieskow, Germany, the place where Hess landed on 7th Aug 1912 after discovery of the ``H\"ohenstrahlung.'' To be published on the {\it Astroparticle Journal}.} 

The problem has been duly mentioned by Vanessa Cirkel-Bartelt \cite{Cirkel-Bartelt:2008fk} in her overview of the history of astroparticle physics taking into account its components \cite[p. 5]{Cirkel-Bartelt:2008fk}. In motivating the use of an historical approach for clarifying the character of the field, she mentioned Todor Stanev having put the question <<Where does the cosmic ray field belong?>>  To which he has given the answer: <<A better definition than an outline of its history and its ever-changing priorities is hardly possible.>>

Even having mainly a biographical character, this outline of Bruno Rossi's scientific activity during thirty years spanning from the end of 1920s to the beginning of 1960s, may provide some  clues to reconstruct  the development of  the strong interdisciplinary, evolutive and creative character of cosmic-ray research, which Rossi himself contributed to shape in the articulated and open form we know today.\footnote{The main biographical source on Bruno Rossi's life and sciences are  his autobiography \cite{Rossi:1990aa}\label{Rossi:1990aa} and George Clark's biographical notes \cite{Clark:2000fk} \cite{Clark:2006aa}.
 Rossi's early work, up to the end of 1930s--beginning of 1940s, has already been discussed by the author in \cite{Bonolis:2011fk} and especially in \cite{Bonolis:2011ak}. For this reason, apart from some aspects which  have not been particularly stressed in these articles, the present work will pay more attention to Rossi's involvement in the beginning of space science and in particular in the birth of X-ray astronomy, as such issues ---strongly related to the main focus of this contribution--- have never been discussed in more detail within the context of Rossi's whole scientific activity, except for descriptions contained in the above mentioned biographical sources. Apart from the scientific literature, the reconstruction is  mainly based on original documents from Rossi's papers preserved at MIT Archives and on research reports of the time, as well as on personal recollections from Rossi's collaborators interviewed by the author.}

\section{Cosmic rays during the 1930s}
\label{sec1}

Born in Venice in 1905, Bruno Rossi studied at the University of Padua, and received his laurea degree in physics at the University of Bologna in November 1927. In early 1928 he got a position at the University of Florence as assistant professor of Antonio Garbasso, the director of the Physics Institute  located on the hill of Arcetri, not far from the house where Galileo Galilei lived part of his life and died in 1642. 

During his first two years in Arcetri, Rossi's research had no clear focus; then a paper by Walther Bothe  and Werner Kolh\"orster  on the nature of the extraterrestrial penetrating radiation appeared \cite{Bothe:1929aa}, which was <<like a flash of light revealing the existence of an unsuspected world, full of mysteries, which no one had yet begun to explore>>  \cite[p. 43]{Rossi:1966aa}.

At the end of 1920s particle physics was on the verge of emerging <<out of the turbulent confluence of  three initially distinct bodies of research: nuclear physics, cosmic-ray studies, and quantum field theory>>  \cite[preface]{Brown-Hoddeson:1983uq}. The  ``radiation from above'' had been studied first as a phenomenon linked to atmospheric electricity, and then as a cosmic and geo-physical phenomenon. During the 1920s the main preoccupation of scientists working on cosmic rays had been to investigate and debate <<the most diverse features of the radiation, such as intensity, distribution, absorption and scattering, and even its origin>> as specified by  Bothe and Kolh\"orster in the opening lines of their pioneering article on the nature of the ``high-altitude radiation'' \cite{Bothe:1929aa}. However, as the authors duly remarked, <<the really essential question regarding the {\it nature} of the high-altitude radiation has hitherto found no experimental answer [italics mine].>> 

At that time it had definitely been clarified  that cosmic radiation had a penetrating power far exceeding that of the $\gamma$-rays emitted by radioactive substances. 
Since $\gamma$-rays were known to ionize through the intermediary of secondary charged particles  which they generate in matter, it was expected that cosmic $\gamma$-rays  traveling through matter would  be accompanied by a flow of  ``soft''  secondary electrons resulting from the Compton effect, which were presumed to be the  ionizing agent recorded by the measuring instruments. 

A direct  study of this corpuscular radiation could thus clarify the nature of cosmic rays. Up to that time, absorption measurement  carried out with highly sensitive ionization chambers had been a main tool for the study of the penetrating radiation. But at the end of the 1920s, brand-new instruments were being developed that opened the door to the investigation of the physical properties of the local radiation, thus starting a real revolution in cosmic ray experimental studies and transforming it into a completely new  research field. A  novel counting tube  had just been developed in Kiel by Geiger and his pupil William M\"uller \cite{Geiger:1928kx} who had announced their invention  on 7 July 1928, during a meeting of the German Physical Society.

In their landmark paper  \cite{Bothe:1929aa} published in the fall of 1929  Bothe and Kolh\"orster reported on their 
measurements of the absorption of those hypothetical ``secondary electrons'' by recording the coincidences between two superimposed Geiger-M\"uller counters interleaved with metal plates of increasing thickness.\footnote{For an in-depth discussion on Bothe's early contributions to the development of the coincidence technique see \cite{Fick:2009fk}; for a review on the birth and development of coincidence methods in cosmic-ray physics see \cite{Bonolis:2011ak}. The topic  was also discussed by Bothe in his Nobel Lecture of December 1954, available  at $\langle$http://nobelprize.org/nobel\_prizes/physics/laureates/1954/bothe-lecture.html$\rangle$.} 
However, since Compton electrons have a low penetrating power,  they should be completely absorbed by a very thin layer of gold. To their great surprise they found that 76\%  of the charged particles present in the cosmic radiation near sea level could penetrate 4.1 centimeters of gold. From this arrangement they argued that  the radiation traversing the two counters within a fixed time coincidences could not have the nature of a secondary radiation resulting from the Compton effect generated by the ultra-$\gamma$-radiation. They thus suggested that the penetrating rays, at that time believed to be of primary origin,  consisted of ionizing particles, and that the ionizing particles observed near sea level were those among the primary particles {\it that were capable of traversing the  atmosphere}. 

Actually, their counter experiments did not really prove that the {\it primary}  
rays were corpuscular, and their latter hypothesis turned out later not to be correct, 
but at the time it had a role in defining the astrophysical and the physical aspects of 
the cosmic-ray problem.  It was still difficult to accept that corpuscles could have the high energies which were necessary to enable them to penetrate the atmosphere.

The pioneering experiments carried out by Bothe and Kolh\"orster during the period 1928--1929 \cite{Bothe:1928vn}  \cite{Bothe:1929lr} \cite{Bothe:1929aa} following the advent of the Geiger-M\"uller counter  in the Spring of 1928, marked the end of the first period of cosmic-ray research. Bruno Rossi in Florence quickly realized that this kind of  physics could be performed in a terrestrial laboratory and that the Geiger-M\"uller counter could become the key to open <<a window upon a new, unknown territory, with unlimited opportunities for explorations>> \cite[p. 35]{Rossi:1981aa}. He was 24 years old when <<one of the most exhilarating periods>> of his life began \cite[p. 10]{Rossi:1990aa}.
In just a few months he built his own Geiger-M\"uller counters, devised a new 
method for recording coincidences \cite{Rossi:1930mz}, and began some experiments. 

Whereas the invention of the Geiger-M\"uller counter  represented the opening of ``a new technological window'', the setting up of the coincidence circuit,  a real {\it telescope} for cosmic rays, constituted a giant leap forward, turning into what it is to be considered the very {\it first act} of Rossi's research program. Rossi's $n$-fold version of the electronic coincidence circuit  changed radically the view of cosmic rays. It greatly reduced the rate of chance coincidences, thus allowing  observations with increased statistical weight.  The time correlation among associated particles crossing different counters could be established, and the possibility of arranging more counters, in whatever geometrical configuration, eventually opened new possibilities of investigation.  

In following Bothe's intuition on the importance of coincidence methods, Rossi's style of work is to be highlighted as exemplary in establishing the ``logic'' research tradition \cite{Galison:1997rm} which paved the way for future investigations. By wisely arranging metal screens and circuits of counters according to different geometrical configurations, Rossi was enacting a definite project which since the very first moment aimed at demonstrating the corpuscular nature of cosmic rays, opposing the view that considered them as high frequency $\gamma$-rays \cite{Rossi:1930qf} \cite{Rossi:1931ul}.\footnote{During the previous years Robert Millikan had provided theoretical justifications  for his theory according to which cosmic rays were the ``birth cry'' of atoms in space, being born in the form of   $\gamma$-rays, from the energy set free in the synthesis of heavier atoms through fusion of primeval hydrogen atoms    \cite{Millikan:1928aa} \cite{Millikan:1928ab} \cite{Millikan:1928lr}\cite{Millikan:1928vn}. For an accurate reconstruction of Millikan's strategy for measuring cosmic ray energy, and his theories on the origin of cosmic rays see \cite[Ch. 3]{Galison:1987kx}   particularly pages 80--89. See also \cite[Ch. 3]{Russo:2000ab}.}

By the summer 1931 Rossi had performed a series of experiments (deviation with an electromagnet \cite{Rossi:1931lq} and study of intensity of cosmic rays with a counter-telescope at different inclinations to the vertical line \cite{Rossi:1931dq}, aiming with different means at studying the nature and behaviour of the cosmic rays.

	Leaving aside the cosmological and planetary scenario of the first two decades of the 20th century, Rossi wondered: <<{\it What is the nature and behavior of this radiation}?>> And in so doing, he was definitely entering the field of microscopic physics. His early researches must be in fact examined on the background of a wider perspective regarding the study of the interaction between radiation, particles and matter, a problem strictly intermingled with the debate over  $\beta$-decay,  behavior of electrons as $\beta$-particles, as well as difficulties in considering the nucleus as a quanto-mechanical system made up of protons and electrons.  Cosmic rays, freely delivered by nature, appeared in fact to be highly relevant  to fundamental problems of physics. 
	
		Rossi discovered that cosmic-ray particles could pass through enormously thick matter, including up to a meter of lead. His findings gave evidence of the astonishing energies associated with cosmic rays. On December 16, 1931, he sent a short letter to \textit{Die Naturwissenschaften} \cite{Rossi:1932cr}: <<A fairly large proportion of the corpuscular radiation found at sea level has still the capability of traversing more than 1 m lead, that is about more than the thickness of the whole atmosphere.>>

 As a matter of fact, the real hot issue was the production of  secondary radiation generated in the metal shield above the counters whose first hints  Rossi had already observed during his experiments in Berlin in summer 1930, when he had been invited by Bothe \cite{Rossi:1931ul}. In fall 1931 Rossi started a systematic research using arrangements of three non aligned GM-counters, in conjunction with his threefold coincidence circuit.  The counters  were placed  inside a thick lead shield in order to avoid any effects caused by external radioactivity. By removing the upper part of the shield the counting rate was significantly reduced, thus affording direct evidence for the occurrence of processes, which resulted in the production of one or more secondary particles by a cosmic ray traversing the matter \cite{Rossi:1932fk}. However, no known process at the time could explain the abundant production of secondary particles revealed by his experiments.\footnote{Such was the novelty of the results, so contrary to the ``common sense'', that the  editors of the scientific journal to which Rossi submitted his short note, most probably \textit{Die Naturwissenschaften},  refused to publish it. It was accepted on February 10, 1932 by {\it Physikalische Zeitschrift}  only after Heisenberg, with whom Rossi corresponded during the period 1930--1932, had vouched for its credibility.\label{rossiheisenberg}}
	
	After the discovery of the production of secondary radiations in metal shields,  Rossi had continued his investigations going much deeper into the origin of this unexpected phenomenon.  Such unpredicted behavior of the radiation found its confirming synthesis in a curve which was to be universally known as the {\it Rossi transition curve}, representing the variation of the number of coincidences recorded by three counters in a triangular array  as a function of the thickness (in mass per unit area) of layers of lead and iron placed above them and emitting the secondary particles \cite{Rossi:1932qy}.

	Rossi's relevant investigations performed during the period 1931--1932 pointed to the existence of two components in cosmic rays
at sea level: a ``hard" component, able to pass through 1\,m of lead
after being filtered through a 10\,cm thick lead screen, and a ``soft" component, generated in the atmosphere by primary cosmic rays and able to
generate groups of particles in a metal screen before being stopped.
Other absorption measurements on cosmic ray particles  had shown the existence of these two components; however, Rossi's experiments were the first to prove that the hard and soft rays   
were fundamentally different in character and did not differ merely in their energy.

 Actually, the Rossi transition curve drops much more rapidly than the absorption curve of cosmic rays at sea level. According to Rossi, this  phenomenon was  due to the fact that what they called at the time ``primary'' radiation, is in part composed of  softer corpuscles of {\it secondary origin} generated  in the atmosphere or in the ceiling of the room and totally absorbed in a few centimeters lead, thus contributing to a steeper slope than expected.\footnote{Actually, since spring-summer 1931 Rossi had introduced a screen of lead 9 cm thick  between two GM-counters (5 cm in diameter and 15 cm long) placed with their axes horizontally, at a distance of 17 cm. By inclining more and more to the vertical line his ``counter telescope'', the slant rays were <<softer and not harder than the vertical ones,>>  contrarily to what  one would expect. The filtering effect of the increasing thickness of the atmosphere should in fact result in a hardening of the revealed radiation. On this occasion he had already stressed that: <<This result may be accounted for by assuming that {\it the corpuscular rays generate in the atmosphere a softer secondary corpuscular radiation} and that the relative amount of the latter is in an inclined direction larger than in the vertical one [\dots]>> \cite{Rossi:1931dq}.\label{slantrays}} 
 
  He was thus able to conclude that :  <<{\it [\dots]  the soft components, and not the hard components of the corpuscular incoming radiation, more actively contribute to the production of secondary radiation.}>>  \cite[p. 256]{Rossi:1932kx}  Rossi remarked that  what he had previously assumed to be a \textit{secondary} radiation generated in metal shields by ``primary'' cosmic rays at sea level, was in fact a <<\textit{tertiary} radiation  producing most of the observed threefold coincidences observed with 1 cm lead>>  and that the soft radiation he had observed <<along strongly slanting directions>> which he considered  <<practically all of secondary origin,>> was at the root of such phenomenon \cite{Rossi:1932uq}.\footnote{Between 1933 and 1934, in order to clarify the nature and origin of the shower-producing radiation, Rossi and his collaborator Sergio De Benedetti performed  a series of absorption and shower experiences at different altitudes (in Asmara and during expeditions to Italian mountains, and in Padua). With increasing altitude the number of coincidences observed with the arrangement of counters out of line increased more rapidly than those with the three counters in line. In the paper ``Una nuova componente della radiazione cosmica''  [``A new component of cosmic-rays''] \cite{Rossi:1934ve}, they hypothesized that <<the shower-producing radiation cannot be considered a secondary radiation of the cosmic corpuscles [the penetrating component]>>  thus concluding that it must be  part of what at the time was considered the ``primary cosmic radiation.''} 
 
 A long paper discussing in detail results on such issues was submitted to the {\it Zeitschrift f\"ur Physik} in February 1933 \cite{Rossi:1933vn}.\footnote{For a discussion of Rossi's early work on the production of secondary showers see \cite{Bonolis:2009fk}. For a more detailed discussion on Rossi's remarkable studies carried out during the period 1930--1932 see \cite{Bonolis:2011ak}} 

In the meantime, after his early attempts in analyzing the corpuscles' reaction in presence of an electromagnet \cite{Rossi:1930ly} \cite{Rossi:1931lq}, Rossi used the Earth magnetic field as a spectrometer \cite{Rossi:1931pd}. If primary cosmic rays were charged particles ---as were showing the above mentioned results--- they would also be affected by the geomagnetic field before entering into the terrestrial atmosphere. A {\it latitude effect} was actually expected indicating a lower intensity of cosmic rays near the equator, where the horizontal component of the geomagnetic field is stronger, but evidence from experiments did not convince the scientific community.\footnote{A slight variation of the intensity had been observed by Jacob Clay since 1927, resulting from experiments made carrying ionization detectors onboard ships that traversed an extensive latitude range stretching from Netherland to Java \cite{Clay:1927fk} \cite{Clay:1928fk}. On the other hand, negative results had been  found by Bothe and Kolh\"orster \cite{Bothe:1930uq}, and by Robert Millikan  and G.H. Cameron \cite{Millikan:1928uq} during recent experiences carried out respectively between Hamburg and Spitzbergen and between Bolivia and Canada.} 
On July 3, 1930 Rossi submitted a paper  \cite{Rossi:1930ve} in which he conjectured the existence of an {\it East-West effect}, a second geomagnetic effect which would be revealed by an  asymmetry of the cosmic-ray intensity with respect to the plane of the geomagnetic meridian,  with more particles coming from East or West, depending  on whether the particle charge was negative or positive. 

After a first negative attempt at Arcetri, Rossi  started planning this  experiment in Eritrea, in the proximity of Asmara, the capital of the Italian colony rising at an altitude of 2370 m, and at a geomagnetic latitude of 11$^{\circ}$ 30' N. In the vicinity of the equator  the effect would be clearly observable.\footnote{Further reflections on the data obtained at Arcetri and on Clay's and others' results, led to a joint article with Enrico Fermi on the theory of geomagnetic effects submitted in February 1933 \cite{Rossi:1933rt}.} Owing to  slow arrival of funds he was able to leave only in the autumn of 1933, thus losing the priority of this important discovery.  When he reported the preliminary positive results of his observation \cite{Rossi:1933fk}, articles confirming his prediction had already been published by Arthur H. Compton and Louis W. Alvarez  \cite{Alvarez:1933yq}, and by Thomas Johnson \cite{Johnson:1933fj}. The study of the asymmetry predicted by Rossi, in confirming  the corpuscular nature of the ``primary radiation'', also revealed that they were mostly positively charged.\footnote{By that time, Arthur H. Compton,  quite impressed by  Rossi's talk at the conference of nuclear physics organized in Rome by Enrico Fermi in fall 1931, had concluded his world-wide campaign confirming and greatly extending Jacob Clay's earlier observations  \cite{Compton:1933vn}. The results showed  that the intensity of cosmic rays is systematically correlated with geomagnetic latitude and  altitude, again confirming their corpuscular nature.} The result was quite surprising because the supporters of the corpuscular theory were convinced, more <<from prejudice than because of a logical argument,>> that the primary particles would turn out to be electrons \cite[p. 36]{Rossi:1990aa}.

As a by product of his experiments in Eritrea Rossi observed an unexpected  phenomenon  \cite[pp. 588--589]{Rossi:1934mz}: 
\begin{quotation}
	\small
	The frequency of the coincidences recorded with the counters at a distance from one another, shown in the tables as ``chance'' coincidences appears to be greater than would have been predicted on the basis of the resolving power of the coincidence circuit. Those observations made us question whether all of these coincidences were actually chance coincidences [\ldots] Since the interference of possible disturbances was ruled out by suitable tests, it seems that once in a while the recording equipment is struck by \textit{very extensive showers} of particles, which cause coincidences between counters, even placed at large distances from one another.\footnote{In the original article Rossi used the Italian words ``estesi'' [extensive] and ``sciami'', a term corresponding to ``swarms'', generally used for a great number of insects like bees, and which since then was used by Italian cosmic-ray physicists. However, the word ``shower'' had just been coined by Blackett and Occhialini, the first to visualize the phenomenon of a spray in fine streams of particle tracks in their cloud chamber triggered by an electronic  circuit.} Unfortunately, I did not have the time to study this phenomenon more closely.
	\end{quotation}
	\normalsize

	At the moment Rossi's remark, which seems to be the first observation of Extensive Air Showers, went unnoticed, mainly because it was included in an article published on an Italian journal. By this time Rossi had many elements which might have connected  the local secondary phenomena generated in matter by the soft component of atmospheric origin --- which he had clearly reported in his articles--- to his observations in Eritrea. However, there is also reason to believe that Rossi felt unsure about the credibility of such results, based on the coincidence between far away counters, which he did not mention in the article published on the {\it Physical Review} \cite{Rossi:1934yq}.\footnote{Actually, during the London Conference of summer 1934, Rossi  stressed that the <<legitimacy of this method has recently been disputed,>> so that he described various experiments with three counters arranged  in different ways in order to <<prove without doubt that the previous results are substantially correct>> \cite[p. 233]{Rossi:1935fk}. And indeed the year before Millikan, who continued to be convinced that the primary rays at sea level and below could not be charged particles, had explicitly attacked the <<so-called Geiger counter coincidence measurements>>: <<I have been pointing out for two years [\dots] \textit{that these counter experiments never in my judgment actually measure the absorption coefficients of anything}.>> Millikan did not  mention Rossi's experiments, which were in fact the main target of his criticism, as Rossi, since the beginning of his activity in cosmic-rays researches, had been the principal proponent of the corpuscular hypothesis.} 
	
	However, it is also to be remarked that at the end of 1933  the complex nature and the mechanisms underlying the whole phenomenon of ``cosmic rays'' was far from being clarified, even if the experiments on geomagnetic effects had definitely provided a proof of their corpuscular nature, and the existence of the positron had just been connected by Patrick M. S. Blackett and Rossi's student Giuseppe Occhialini to Dirac's theory of the relativistic electron and to the mechanism of pair production.

Patrick M. Blackett and  Occhialini \cite{Blackett:1932mz} had in fact used Rossi's coincidence circuit to trigger the expansion of a cloud chamber. With this new device, a perfect fusion of  ``image'' and ``logic'' \cite{Galison:1997rm}, they were able to ``see''  the groups of particles whose existence Rossi had clearly demonstrated in his experiments as  some sort of a ``rain'' in the cloud-chamber, hence they named them ``showers''. In their second paper \cite{Blackett:1933fj}, Blackett and Occhialini pointed out that the occurrence of these tracks was a well known feature of cosmic radiation and was <<clearly related to the various secondary processes occurring when penetrating radiation passes through matter.>> They also credited Rossi for having been the first to investigate these secondary particles using counters and were able to confirm the interpretation proposed by Anderson about the existence of a <<positively charged particle having a mass comparable with that of an electron>> \cite{Anderson:1932fk}, but they also placed the positron  ---the ``Dirac hole-theory particle''--- into theoretical perspective. Blackett and Occhialini were the first to expound the pair formation mechanism derived from experiments, which was the process underlying the formation of electromagnetic showers, one of the most striking facts of the phenomenology related to cosmic rays.\footnote{In 1933 Bhabha had written his first article on cosmic rays identifying Rossi's results obtained with three non-aligned counters with the showers observed by Blackett and Occhialini. Even if he was writing in German, he used the English word  ``Showers''  they had just ``invented''  \cite[p. 120]{Bhabha:1933fj}.} But only starting from 1934, with the Bethe-Heitler's  landmark paper ``On the Stopping of Fast Particles and on the Creation of Positive Electrons''  \cite{Bethe:1934fj}, discussing in detail the energy loss of charged particles passing through matter based on the Dirac equation, views on cosmic-ray phenomena really began to change. 

In the period when understanding of the interaction between cosmic rays and matter was far from clear and quantum electrodynamics appeared to break down at the high energies involved in cosmic rays, experiments using counters became one of the fundamental tools for testing the new physics. A theory providing a natural and simple explanation on the basis of the quantum electrodynamics cross-sections calculated by Bethe and Heitler   appeared in early 1937 in simultaneous papers by J.\ Franklin Carlson and J.\ Robert Oppenheimer \cite{Carlson:1937kx}, and by Homi Bhabha and Walter Heitler \cite{Bhabha:1937yq}. The theory explaining shower formation and the identification of the hard component with a brand-new particle, the mesotron, resurrected relativistic quantum field theory toward the end of the 1930s \cite{Galison:1983fk} \cite{Brown:1991fk}.\footnote{The logic tradition was instrumental in dealing 
with large samples of events connected to a single kind of phenomenon.  And indeed, Bhabha and Heitler used the ``Rossi transition curve''  to establish the correctness of their theoretical results \cite{Bhabha:1937yq}.}
The core of the theory was the explanation of the activity regarding the soft part of cosmic rays, the unequivocal success of quantum electrodynamics, which nevertheless left unsolved the study of the hard component leaving it as a challenge for the future.

In parallel with theoretical clarifying of the shower formation, which clearly connected creation of showers in matter to the existence of similar phenomena in the atmosphere,\footnote{Bhabha and Heitler concluded their article making a comparison between Rossi's transition curve and Regener's absorption curve in the atmosphere, finding that both could be understood on their theory \cite[p. 458]{Bhabha:1937yq}.} Georg Pfotzer  had concluded that soft rays observed at low level were to be considered as branches of showers produced high in the atmosphere by incoming cosmic-ray particles \cite{Pfotzer:1936fk}.\footnote{In clearly talking about  primaries and secondary components, Pfotzer identified the soft component with electrons and positrons connecting it with showers (he used the word ``Schauer''), and the hard component with protons, producing most of the effects at sea level.}
 These opinions were confirmed in the concluding section an article published by Pierre Auger and Paul Ehrenfest jr. in 1937 \cite[p. 206]{Auger:1937fk}. In starting their study on the two components of different nature discovered by Rossi, Auger and his collaborators made systematical researches on the   showers produced by soft cosmic rays in matter.\footnote{Actually Auger had been involved in such researches by Louis Leprince-Ringuet, who, in turn, had been convinced by Rossi in 1932 about the interest of cosmic-ray studies and of electronic coincidence methods.} They had been led to extend the study to distances of several tens of meters, progressively arranging counters more and more wide apart in order to reveal more and more dispersed atmospheric showers.  They explored the production of large atmospheric showers (<<grandes gerbes cosmiques>>), whose branches could be separated by a distance of several meters , and whose origin they thought should be <<very high in the atmosphere>> \cite[p. 1721]{Auger:1938rt}. In comparing such showers to those <<at small angles>> described by Schmeiser and Bothe \cite{Schmeiser:1938fk}, they remarked how their existence confirmed that many  particles are formed at sea level in secondary multiple processes generated in the atmosphere \cite[p. 1723]{Auger:1938rt}. Auger and collaborators, found that such showers could cover a surface of the order of 1000 m$^{2}$, implying the presence of several dozens of thousands of particles  and involving energies bigger than 10$^{13}$ eV, probably <<beyond  10$^{14}$>> \cite[p. 229]{Auger:1938fr}. In that same 1938, Kolh\"orster and collaborators reported about coincidences recorded between two counters  separated by distances going up  to 10 m, and remarked that these observations must be related to <<secondary rays of the high-altitude radiation, i.e. to a shower [Schauer]>> \cite{Kolhorster:1938uq}.\footnote{For a comparison between Auger's, Bothe and Schmeiser's, and Kolh\"orster's experiments see  Figure 3 on K.-H. Kampert  and A. A. Watson's historical review on EAS and ultra high-energy  cosmic rays  \cite{Kampert:2012fk}.}

A summary of the French group systematic researches on the extensive air showers appeared in an article published on the {\it Reviews of Modern Physics}, where they discussed the extension of the energy spectrum of cosmic rays up to  the incredible energy of 10$^{15}$ eV. One of the consequences was actually that it appeared <<actually impossible to imagine a single process able to give to a particle such an energy>>  \cite{Auger:1939aa}. Auger and his collaborators were since then credited for the discovery of EAS, and all this inaugurated the great season of speculations on the origin of cosmic rays which is still an open question nowadays.

\section{The mesotron decay}
\label{sec2}

Supported by Enrico Fermi, Rossi had won a competition for the chair of Experimental Physics and worked in Padua from 1932 to 1938. At this time, after having just completed his work overseeing the design and construction of the new Physics Institute, he was rewarded by being denied the right to work there because he was a Jew. When the anti-semitic laws enacted by the fascist regime of Benito Mussolini passed in Italy, Rossi was dismissed and had to emigrate from his country with his young bride Nora Lombroso.\footnote{A detailed reconstruction of Rossi's forced emigration from Italy can be found in  \cite{Bonolis:2011fk}.}

They left Italy in September 1938. After a short stay in Copenhagen at Niels Bohr's Institute, Rossi was invited by Patrick Blackett  in Manchester, and in June 1939 he moved to the United States on an invitation of  Arthur H. Compton.\footnote{Immediately after his arrival in the USA, Rossi became  more and more involved in early researches regarding cosmic-rays as a source of  ``elementary particles''. Being  considered a leading figure in the field, he was asked to prepare  reviews on cosmic-ray problems of the moment  \cite{Rossi:1939zl} \cite{Rossi:1939fk} \cite{Rossi:1940pd} \cite{Rossi:1941la}.}  

 In the period 1939--1942 Rossi conducted with different collaborators a series of experiments  on the decay of the first unstable particle which had been discovered in cosmic rays in 1936--1937, the $\mu$ meson, or ``mesotron''  of cosmic rays, as this particle was then called. These experiments  produced the final proof of the radioactive instability of the mesotron, established a precise value for its mean life, and, as a by-product, verified for the very first time the time dilation of {\it moving clocks} predicted by Einstein relativity theory \cite{Rossi:1940gf} \cite{Rossi:1941wd} \cite{Rossi:1942ai} \cite{Nereson:1943bh}. By 1942 the evidence for the decay of the mesotron at the end of their range had changed from the first two cloud-chamber tracks photographed by Williams and Roberts in 1940 \cite{Williams:1940rt} to the curve presented by Rossi and Nereson, which contained thousands of decay events and showed an exponential decay with a lifetime of about 2 $\mu$s  \cite{Nereson:1943bh}. The style and elegance of these achievements have been unanimously recognized  in the history of experimental physics. 
 
 These experiments, completed by Rossi during the war, symbolically closed an era that he himself called <<the age of innocence of the experimental physics of elementary particles>> \cite[p. 204]{Rossi:1983uq}. It was still a time when fundamental results  could be obtained through extraordinarily simple experiments, which cost a few thousand dollars, and which required the help of one or two young graduates. 
 
 The years from 1939 to 1943 form a self-contained period of Rossi's personal life and of his scientific activity. It began with his arrival in the United States, as an exile from fascism, and ended with his shifting from the peace-time work at Cornell University to the war-time work at Los Alamos. From the late spring of 1943, he participated to the Manhattan Project at Los Alamos  where the best brains were gathered to work on research aiming at the production of nuclear weapons. With Hans Staub he was responsible for development of detectors for nuclear experiments. Their \textit{fast ionization chamber} was used in a series of tests conducted by Rossi that verified  the implosion method for detonating the plutonium bomb  and in a measurement he made of the exponential growth of the chain reaction in the Trinity Test,  the first ever nuclear explosion, on July 16, 1945.

\section{The invasion of accelerators}

The importance Rossi had acquired within the American scientific context can also be measured by the offer of a position at the Massachusetts Institute of Technology at the end of the war. He moved there in early 1946 and established the Cosmic Ray Group  at the Laboratory for Nuclear Science and Engineering.\footnote{In the Academic Personnel list of the Physics Department illustrating the different members of the MIT faculty concerned with the operation of the program in Nuclear Science and Engineering, Rossi was presented as <<one of the two or three most distinguished experts in the country in cosmic rays.>>}
 Excited by the return to the active research and by the possibility to form a new generation of researchers, he  started to think on a grand  scale, relying on the  large economic resources which physicists had at their disposal after the scientific and technological achievements of war research. 

Modern elementary particle physics as such, was on the verge of becoming an independent field, but still strongly depending on cosmic ray research, whose instruments consisted largely of cloud chambers, ionization chambers, nuclear emulsions and counters, of course. 
 
An impressive research program aimed to study the production of mesons,  their interaction with matter, and their decay process by all possible means, included high-altitude mountain experiments, airplane measurements and the study of giant cosmic ray showers of the atmosphere, which would become one of the  MIT Cosmic-Ray Group's strongest research fields. At the same time, the problem of the identity of primary cosmic rays was tackled by means of suitable ionization chambers and amplifiers, carried by balloons. This vast and complex program on cosmic rays and elementary particles saw Rossi surrounded by an increasing number of young researchers, often coming from abroad to enrich the cultural circle within the local scientific community.

Up to the middle of the 1950s the ``little science'' program carried out by the cosmic ray physicists with counters, cloud chambers and nuclear emulsions, continued to be competitive with ---and even to dominate--- the ``big science'' of particle physics carried out using accelerators. But since some time Rossi's interest in the primary cosmic radiation had been aroused by experiments carried out toward the end of the 1940s. At that time, the secondary processes occurring in the atmosphere and generating the various components of the local radiation (penetrating component, shower-producing component, nuclear-active component) had been clarified, but cosmic-ray research was undergoing a radical change. 

On the one hand, in 1940 balloon experiments by Schein and co-workers \cite {Schein:1941zr}    using complex Geiger-M\"uller counter arrangements, and in 1948 balloon experiments using nuclear emulsions \cite {Bradt:1948aa}  \cite {Bradt:1948fk},  had solved the problem concerning the nature of the primary cosmic radiation, three decades after Hess' seminal discovery. By 1950 the main features of the composition of primary cosmic rays were thus known. 

On the other hand, since the 1930s the potential importance of cosmic rays for astrophysics had been pointed out.\footnote{In 1934  Baade and Zwicky had related the appearance of supernovae to the formation of neutron stars and the generation of cosmic rays \cite{Baade:1934vn}.} Research on the primaries  began thus to stir up questions connected with the origin of cosmic rays, but at the moment problems related to nuclear interactions  were still overwhelming.

 In fact, as recalled by Rossi, the discovery of the strange particles and the attempts <<to establish their nature and behavior became one of the most fascinating and demanding activities of our group>> \cite [p. 117]{Rossi:1990aa}. At that time the activities of the MIT Group were in fact divided between experiments on cosmic rays and experiments with accelerators. At the beginning of the 1950s a semblance of order began to appear, but  a much deeper understanding of nuclear forces still needed a lot of ingenuity both by theorists, and by accelerator experimenters and cosmic-ray physicists. In this period, the MIT Cosmic Ray group and the \'Ecole Politechnique  group led by Louis Leprince-Ringuet had a leading role in resolving the charged kaon group. 

 This work was part of a series of  studies of high-energy cosmic-ray events in multiplate cloud chambers, in the course of which stopping kaons were found \cite{Rossi:1952ay} \cite{Rossi:1953dp} \cite{Rossi:1953mw} \cite{Rossi:1953rw} \cite{Rossi:1953wj} \cite{Rossi:1954fx}. 
 
 As remarked by Robert Marshak, by the time of Rochester III (Annual Conference on high-energy Nuclear Physics), held 18--20 December 1952  \cite {Marshak:1989aa}:\footnote{Rossi presided the Experimental Physics Session, whose results were both coming from cosmic-rays and accelerators.} 
 \begin{quotation}
 \small
 The machine results were certainly overtaking the cosmic ray results in pion physics and the theorists were frantically trying to develop a plausible theory of the pion-nucleon interaction starting with isospin invariance. But, absent the completion of the 3 GeV Cosmotron at Brookhaven, the cosmic ray physicists still dominated strange particle physics.
 \end{quotation}
 \normalsize

 Those years would culminate in the 3rd International Cosmic Ray Conference, the great congress held at Bagn\`eres de Bigorre in 1953, which old people still remember with nostalgia. Not only was a major occasion of encounter for the groups involved in the study of elementary particle physics all over the world, but also provided the opportunity for examining the data collected in the previous years in an attempt to clarify a very confused situation in the field.\footnote{For a reconstruction of the story of such remarkable conference see \cite{Cronin:2011qy}.} 
 Cosmic ray research and high-energy particle physics were still sharing a remarkable connection, even if the feeling that accelerators were beginning to invade the field began to spread during the Bagn\`{e}res-de-Bigorre Conference.  As recalled by Milla Baldo Ceolin, at the time a young researcher of Padua University, <<order began to emerge from chaos>> \cite [p. 7]{Baldo-Ceolin:2002aa}.  Milla Baldo Ceolin well remembered how Rossi dominated the scene at Bagn\`ere de Bigorre.\footnote{Interview of the author with Milla Baldo Ceolin,  November 10, 2006. Rossi's presence was considered fundamental by  Leprince-Ringuet, secretary of the conference, who had written him on October 20, 1952 asking for suggestions and remarking that the success of the event would mainly rely on his participation. L. Leprince-Ringuet to B. Rossi, Rossi Papers, MIT Archives, Box 19, Folder ``General Correspondence June 1952--November 1953'', Subfolder Sett.--Dec.1/2.}

   In October 1954 the G-stack, a giant stack of 250 sheets of emulsions was prepared by a European collaboration. It had a dimension of $37\times27$ cm$^{2}$ in order to follow decay products to the end of their ranges, flew for six hours at 27,000 meters of altitude. The cosmic ray physicists could be proud; they had found just in time all possible decays of the heavy mesons, and made it very plausible that there was one and only one $K$ particle. The ``wisdom'' and ingenuity of Rossi's talk at Bagn\`eres de Bigorre received a confirmation at the Pisa Conference in July 1955. The G-stack experiment showed that the $K$-mesons, having different decay modes, had masses that were equal within a few tenths of 1\%.  The last great experiment of cosmic rays in particle physics was signed by 36 authors instead of the few typical of those times.\footnote{The results of the launch, flowing 93 kg of emulsion stacks, were measured and analyzed by a collaboration formed by a large number  of laboratories, twenty-one European and one Australian. For a detailed historical reconstruction of the experiment see \cite{Olivotto:2009fk}}
   
 But their triumph was a swan's song.  In the meantime the new generation of machines had begun to produce results comparable to those obtained by cosmic ray physicists. 
 In 1953 the first strange particles produced in laboratory experiments using accelerator beams were observed. 
 It was just the beginning.
  
  In June 1952 Rossi had completed a most arduous task of preparing the volume \textit{High-Energy Particles} \cite {Rossi:1952aa}, a full survey of the information gathered up to 1951 about the elementary particles.\footnote{In what was an advanced review textbook, as well as a precious source of references, Rossi presented the most important experimental data, together with many theoretical developments. Over 500 references were quoted, two-thirds of which had been published since 1945. For many years Rossi's book was  a ``Bible'', and a constant companion for high-energy physicists.}  
 By that time the study of the complex phenomena associated with elementary particles was increasingly pursued for its own sake, achieving independence as a new branch of science called high-energy physics. Puzzles and uncertainties  about the mysterious decay processes of strange particles began to receive  a theoretical explanation. In the words of Milla Baldo Ceolin  \cite [p. 17]{Baldo-Ceolin:2002aa}:
 \begin{quotation}
 \small
The September 1957 Padova-Venice International Conference on ``Mesons and Recently Discovered Particles'' was characterized by a completely new climate. It was the year of great theoretical success, when the breakdown of parity conservation was confirmed by experiment, the antiproton and the neutrino had been discovered, the $K^{0}_{2}$ detected, neutrino oscillations predicted [\dots] accelerators had replaced cosmic rays as the principal source of high-energy particles. The stream or results from the Cosmotron and the Bevatron monopolized strange-particle physics, while bubble chambers, counters, and spark chambers were slowly replacing cloud chambers and nuclear emulsions as the principal detectors [\dots]  
 \end{quotation}
 \normalsize

  From the mid-1950s, the interest in cosmic rays shown by high-energy physics fell sharply and cosmic ray research underwent an evolution which gradually changed its old character.   The hunt for elementary particles was an exciting adventure and some  cosmic ray physicists  turned to work with accelerators. But the cosmical side of their research had been always alive in the heart of  those whose interest became increasingly focussed toward the astrophysical aspects of the cosmic-ray problem.
Already in 1949, in presenting the status of cosmic ray research,  Rossi had well expressed the double nature of the identity of cosmic-ray physicists:\footnote{B. Rossi, ``Present status of Cosmic Rays'', draft of a talk, Rossi Papers, MIT Archives, Box 22, Folder ``1949''.}

\begin{quotation}
\small
[\ldots] one can look at cosmic rays as an {\it astronomer} or as a {\it physicist} [emphasis added].

If you look at cosmic rays as an astronomer, you are likely to ask yourself questions like: What and how are cosmic rays produced? How are they distributed throughout the universe? How large is their energy, confronted with other forms of energy in the universe? What is the influence on cosmic rays of the magnetic field of the galaxy, of the Sun, of that of the Earth? Can we learn anything about the structure of these fields from the study of cosmic rays?

If you look at cosmic rays with the eyes of the physicist, you mainly see in the cosmic radiation a source of very high-energy rays [\dots]   that you  can use to obtain information on the properties of elementary particles on the one hand and on the properties of matter, and more particularly of atomic nuclei, on the other hand.

\end{quotation}
\normalsize

At that time cosmic rays were still mainly considered as a tool for nuclear research, and the first question to answer was: What kind of particles are cosmic rays made of? However, Rossi also stressed that the astrophysical problem was a <<most fascinating>>  one\dots 

By 1953, when it had become definitely clear that the future of high-energy physics lay in the accelerator laboratory, the interest in cosmic rays definitely shifted to the problems of their origin from still unknown sources and propagation in astrophysical environments. From then on, scientists like Rossi and others were definitely attracted towards phenomena taking place in space outside the Earth.

\section{Outer space: Extensive Air Showers and all that}

Already Auger and his collaborators had showed in the late 1930s that there exist in nature particles with an energy of $10^{15}$ eV \cite{Auger:1938fr} \cite{Auger:1938rt}  \cite {Auger:1939aa}, at a time when the largest energies involved in natural radioactivity processes were just a few MeV. At its completion in 1952, the Brookhaven Cosmotron was the first to reach and pass $10^{9}$ eV energies, but  by that time new and old techniques were opening the way to new approaches able to determine the energy of primary cosmic rays which had been studied since the end of the war with the use of rockets, air balloons and aircrafts disposed by the Navy. At  the end of the 1940s remarkable results had provided hints of the existence of  primary particles having energies in the range up to $10^{17}$ eV. The hunt for such high energy events had been a strong motivation for a specific air shower research project carried out by Rossi's group at MIT since early 1946.

 The conclusions reached by Auger and his colleagues depended on the statistical analysis of many events, aided by the newly developed theory of electron-photon cascades, while this kind of experiments required detectors whose signals, unlike those of the GM-counters,  would be proportional to the numbers of shower particles which had struck them. Advances in the experimental technique using of the pulse ionization chamber as an instrument of high sensitivity and microsecond time resolution, made it possible to investigate in greater detail the structure of these air showers also enlarging the detection area.

Because of scattering, particles of a shower can be found at large distances from the axis. A sampling technique can thus be used which was  applied for the first time by Robert W. Williams, Rossi's collaborator at MIT, who established an array of 4 ionization chambers to locate the position of the central region of the shower. 
 In the late 1940s, these experiments, in showing how to find individual core positions from sample of particle density, for the first time proved conclusively the existence of showers with more than $10^{6}$ particles  \cite {Williams:1948fk}. This was a first step toward finding energies and directions of individual showers.\footnote{From July  to the middle of September 1951, a newly rebuilt multiple ionization chamber array was operated at Echo Lake. This equipment was used to trigger the expansion of a large cloud chamber located near the center of the array. More than 1000 events initiated by particles of $10^{14}$ eV were recorded, and it was hoped that the analysis of ion-chamber pulse heights might clarify the structure of air showers near the shower core.}

In the meantime, the technique of scintillation counters was being developed, and it became clear that these would be more suitable detectors for air shower experiments than ionization chambers. During the summer of 1952 Rossi suggested that George W. Clark  and Piero Bassi, visiting from the University of Padua, look into the possibility of determining the arrival directions of extensive air showers by measuring the relative arrival times of shower particles over an array of scintillation detectors. Their experiment, with three detectors on the roof of an MIT building, validated the fast-timing method for determining the arrival directions of air showers \cite{Rossi:1953rm}.\footnote{For a reconstruction of the history of MIT Air Shower project see    \cite{Clark:1985aa}.} 

However, dangers connected with the fire hazard coming from liquid detectors resulted in the development of large plastic scintillators, an important technical advance which between May 1956 and spring 1957 allowed recording of showers with more than $10^{8}$ particles at the rate of about one per month. The largest air shower recorded up to that moment had 1.4$\times10^{9}$ particles and the energy of the primary particle responsible for this shower was estimated to be close to $10^{19}$ eV.\footnote{They estimated that it could not have been smaller than 2$\times10^{18}$ eV, even allowing for a large fluctuation in the longitudinal development of the shower. Laboratory of Nuclear Science\&Engineering (from now on LNS\&E), \textit{Annual Progress Report}, May 31, 1957, p. 64.}
With this event, the MIT Group was extending the energy spectrum of primary cosmic rays.  

The fundamental role played by this ambitious research program in the general frame of Rossi's interest towards the astrophysical aspects, has been recalled by George Clark. He remarked how their primary purpose <<was to cast new light on the problem of the origin of cosmic rays by a determination of the primary energy spectrum and distribution of arrival directions of primaries with energies above $10^{14}$ eV>>  \cite [p. 240]{Clark:1985aa}. The transition to a new stage of Rossi's scientific life is testified by his keen interest in the origin of cosmic rays which was becoming a hot problem also because, on the other hand, cosmic ray physicists had now the possibility of studying the composition, the energy spectrum, the time variations and the directions of arrival of primary cosmic rays in an effort to get some clues on their place of origin, on the mechanism responsible for their production, and on the conditions prevailing in the space through which they travel on their way to the earth.\footnote{The titles of international  Summer Schools organized in 1953 and 1954 at  Varenna, on the Como Lake,  well express the changing times and the fast evolution of research.  During the second Varenna course of 1954 Rossi held two lectures---``Fundamental Particles'' and  ``Origin of Cosmic Rays'' \cite{Rossi:1955fk}  \cite{Rossi:1955lr}--- whose titles testify the ongoing evolution of his identity as a cosmic-ray physicist.} 

The discovery of the existence of particles characterized by energies much  higher than could be previously imagined, had revived the interest for mechanisms able to accelerate cosmic rays at such energies. In 1949 Fermi  had published a most cited article  \cite{Fermi:1949dq} where he considered cosmic rays as a gas of relativistic particles moving in the interstellar magnetic fields. He proposed for the first time  a model for the acceleration mechanism responsible for the production of cosmic rays which could be compared with data. 

Between 1953 and 1954 the origin of cosmic rays had definitely become a hot topic and various theories were published \cite {Rosen:1969aa}. Fermi  got interested in magnetic galactic fields during discussions with Hannes  Alfv\'en and especially with Subrahmanyan Chandrasekhar, and  took the occasion to re-examine his earlier ideas about the origin of the cosmic rays in view of later developments in the knowledge of the strength and behavior of the magnetic fields \cite{Fermi:1953aa}  \cite{Fermi:1954aa}. At the time Rossi, Philip Morrison and Stanislaw  Olbert were writing an article on the origin of cosmic rays \cite{Rossi:1954bd}, too, so that the issue was discussed in a correspondence  which was forcibly interrupted with Fermi's premature death in the late autumn of 1954.\footnote{Fermi's paper appeared in January on \textit{The Astrophysical Journal}, while Rossi's article was published on the April 15 issue of {\it The Physical Review}. Both were included  in a  volume  of 76 selected papers on {\it Cosmic Ray Origin Theories} (ranging from 1932 to 1966) \cite {Rosen:1969aa}, appearing  one after the other in the summary,  a nearness symbolizing their longstanding dialogue on cosmic rays as well as their everlasting scientific and human relationship.} 

The study of the origin of cosmic rays, in parallel with the interpretation of empirical evidence obtained from properties of the cosmic rays themselves, began to draw as well from general astrophysical means of observation. The beginning of the 1950s assisted indeed at the birth of cosmic-ray astrophysics. The connection of primary cosmic rays with  chemical and isotopic compositions of matter in the universe, with energy spectra of protons and nuclei, their origin and the mechanisms for their accelerations and the propagation in the interstellar medium, were all problems which provided a foundation for the rising of the concept of cosmic rays as ``astronomical objects'' along with galaxies, stars, interstellar gas etc. \cite {Ginzburg:1988fk}.

The development of more and more sophisticated air showers experiments, in connection with the ongoing evolution of the cosmic-ray problem,  certainly represented a strong contribution in  redirecting Rossi's interest towards phenomena specifically connected to the ``cosmic character'' of the radiation. A whole era was concluding in his personal line of research, which had opened with his first efforts in providing the proofs of the corpuscular nature of the radiation, until the unveiling  of the rich, but still rather confused, universe of particle physics during the 1940s and early 1950s. 

Magnetic fields, particles generated in the explosion of supernovae, and nuclei of elements expelled by astrophysical objects of unknown nature were more and more crowding  the intergalactic space. At that time only physicists were interested in cosmic rays, their role in astronomy was still completely obscure, mostly because of the high degree of isotropy of the cosmic rays (ignoring the effect of the magnetic field of the Earth). For this reason even the very detailed information available on the composition and the energy spectrum of the cosmic rays on Earth said little about the nature ---and especially about the very location--- of these sources. However, by the early 1950s the synchrotron nature of a significant part of the cosmic radio emission was established and it became possible to obtain information on the electron component of cosmic rays far from the Earth, in the Galaxy and beyond its limits as well. The establishment of the relationship between radio emission by cosmic sources and the physics of cosmic rays was leading to the birth of the astrophysics of cosmic rays and then to high-energy astrophysics, thus shaping a new path for cosmic-ray research.

 Since 1957 a decision had been reached by Rossi's group at MIT about extending the air shower investigations to allow for detection of higher energy events.\footnote{It was expected that primaries with energies near $10^{20}$ eV would be observed (<<if they exist>>), LNS\&E \textit{Annual Progress Report}, May 31, 1958, p. 65. For this experiment, a special grant of \$150,000 was given by the National Science Foundation.}   
 The Volcano Ranch experiment, the first giant array with an area of 2.2 km$^{2}$, was constructed by John Linsley with the collaboration of Livio Scarsi  in New Mexico. It led  to the discovery of primary particles having the astonishing energy of  $10^{18}$ eV and even of more than $10^{19}$ eV, at the time the largest ever registered \cite{Rossi:1961yq}.

After Linsley enlarged the array to an area of 10 km$^{2}$, the experiment established a new record of shower size. The largest showers recorded contained 5$\times10^{10}$ particles, with a corresponding estimated energy of the primary particle responsible for its production of 10$^{20}$ eV. The primary spectrum as obtained from   data collected by the MIT group up to that moment showed an extension without a break for the the primary spectrum from several times 10$^{15}$ eV to about 10$^{20}$ eV. Such  energies extended measurements of the primary spectrum well beyond the range of possible galactic confinement. These results, of basic importance for the problem of the origin of cosmic rays, were first reported at the International Conference on Cosmic Rays and the Earth Storm held 5--15 September, 1961, in Kyoto, Japan  \cite {Linsley:1962yq}. 

It was clear that the discovery of  ultra-high energy primaries was a precious reward for physicists who had  accepted the challenge of searching for such rare events, as  Rossi himself well expressed in 1959:\footnote{B. Rossi, ``Cosmic Ray Particles of Extreme Energies (Rough Draft)'', N.Y. Lectures -- April 1959. Rossi Papers, MIT Archives, Box 13, Folder  ``Lectures and Notes 1959''.}

\begin{quotation}
\small
Before the mounting pressure of man-made accelerators, some of us cosmic ray physicists have retreated to the high-energy end of the cosmic-ray spectrum = the safe region above the energy of $10^{15}$ eV that would be produced by a bevatron surrounding the whole earth.
\end{quotation}
\normalsize

In entering the realm of the ultra-high energy universe, cosmic ray physicists were  definitely changing their identity,  having already avoided the ``risk'' of  getting lost in what Beppo Occhialini called the <<monocoltural world of particle physics.>> 

As Livio Scarsi remarked in 2004, not long before his passing: <<Today the search of the Extreme Energy end of the Cosmic Radiation spectrum is one the most exciting chapters in Astroparticle physics.>> \cite [p. 6544]{Scarsi:2005aa}. 

\section{Physics in Space}

A full-scale political and cultural crisis resulted from the launching on October 4, 1957, of {\it Sputnik 1}, the world's first artificial satellite and Soviet Union's contribution to the International Geophysical Year. 

On January 31, 1958, the first U.S. satellite, {\it Explorer I}, went into orbit carrying a GM-counter designed by James Van Allen's group. It discovered the great belt of trapped radiation surrounding the Earth thus deserving the credit for the first important scientific discovery of space exploration. 

During the months preceding the passage of the National Aeronautics and Space Act of 1958 scientist <<were united in their desire to have a strong scientific component in the space program.>> James Killian, President of the Massachusetts Institute of Technology, and special assistant to the President for science and technology, as well as the other Members of the President's Science Advisory Committee (PSAC) <<saw in the space program an opportunity to renew national support of science.>> These groups and the members of the Academy of Sciences pressed for a space program in the direction of science and under civilian management \cite[Ch. 7]{Newell:1980aa}.

In the summer of 1958  Dwight D. Eisenhower signed an act establishing the National Aeronautics and Space Administration (NASA) and at the same time, the National Academy of Sciences created a \textit{Space Science Board} (SSB) to interest scientists in space research and to advise NASA and the other federal agencies the Academy expected to be engaged in space research. Bruno Rossi, who was among the fifteen members of the \textit{Board} called together for their first meeting in New York on June 27, 1958, was asked to form  a Committee on special space projects.\footnote{Minutes of the first meeting Space Science Board, June 27, 1958. Rossi asked Thomas Gold, Salvador Luria and Phillip Morrison to work with him in the Committee.} During the discussion by each committee chairmen addressing to fundamental questions to be answered  regarding future experiments, Rossi focused on <<Gamma ray astronomy -- gravitational red-shifts. Long-range visionary experiments and/or programs.>>

As a reaction to the growing importance of the ``cosmic dimension'' in their researches Bruno Rossi organized a series of conferences and seminars devoted to space-related topics, as he wrote at the beginning of 1958 to Herbert Bridge, one of his most important collaborators, on leave at CERN:\footnote{B. Rossi to H. Bridge, December 26, 1957, Rossi Papers, MIT Archives, Box 23, Folder ``Back Correspondence''.}
 \begin{quotation}
\small
The new development here is a rapidly increasing interest in what we are presently calling ``cosmic physics'', which is of course a natural extension of our interest in cosmic ray proper. Next term we are starting a seminar in which we are planning to discuss some of the problems of the primary cosmic radiation along with problems of magneto-hydrodynamics, radio astronomy, etc. 

\end{quotation}
\normalsize

By this time it turned out to be very natural for the Cosmic-Ray Group to redirect a main part of their activities toward the new space-oriented program, having opportunities for this kind of experiments become available. Only Linsley and Scarsi remained assigned to the ``ready to go'' Extensive Airs Showers Volcano Ranch Project \cite[pp. 153--154]{Scarsi:2007fk}.

However, in Rossi's and his collaborators' opinion, space vehicles were not to be used for the purpose of obtaining better information on primary cosmic rays than it was possible to do from earth-bound stations. Rather, they had clear that space experiments had to be devised that would provide a \textit{new} line of attack to the same astrophysical problems in which they had become interested through cosmic-ray studies. This led into two research programs, both related to cosmic-ray research proper: interplanetary plasma and $\gamma$-ray astronomy.

 Bruno Rossi did not directly participate to the MIT satellite $\gamma$-ray astronomy\index{gamma-ray astronomy}  which was initiated by William L. Kraushaar in 1958, but he immediately promoted the project as a member of the {\it Space Science Board}. George W. Clark joined W. L. Kraushaar, and both designed and built a $\gamma$-ray telescope at MIT's Laboratory for Nuclear Science and directed the experiment which led to the first observations of high-energy ($>$ 50 MeV) cosmic rays  \cite{Kraushaar:1962qa}  \cite{Clark:1962qy}  \cite{Clark:1965uq}.  

The discovery of the Van Allen radiation belts in spring 1958 had been made using a very simple and basic instrument  such as the the old well known GM-counter. All this was music to cosmic-ray physicists' ears. They had a close association with this glorious device since the beginning of their scientific enterprise, moreover the result demonstrated, too,  that  a university physics department could design and build instruments capable of operating in space and producing significant scientific results. Actually, it was not by chance that  in early space age this happened  thanks to cosmic ray physicists.

Observations from the Earth had already suggested that the space surrounding our planet is not entirely devoid of matter, as had been supposed in the past, but rather contains a dilute plasma,  an ionized gas that was considered presumably consisting of electrons and protons. The Cosmic-Ray Group at MIT had been concerned with this problem because of the possibility suggested by some scientists that certain temporal changes of cosmic-ray intensity might be due to clouds of magnetized plasma ejected by the Sun into the surrounding space. At the end of the 1950s most scientists were thus persuaded that a plasma of solar origin was filling interplanetary space all the way to the Earth's orbit and probably beyond.  However, the views concerning the properties of this plasma were widely divergent. The estimates of its density ranged from one to one thousand electrons and protons per cubic centimeter. Some believed that the plasma was nearly stationary, others that it was flowing with a speed of 1000 kilometers per second. Perhaps the plasma was distributed more or less evenly in space; perhaps it was condensed into clouds.
 
   In parallel with the analysis of observational data, there  existed two pictures of interplanetary space. Ludwig Biermann's stream of particles flowing out from the Sun at high speeds,\footnote{In 1951 Ludwig Biermann interpreted  deflection of cometary gas tails in terms of the interaction between the cometary ions in the tail and a stream of particles with densities for the quiescent Sun of the order of some hundreds particles per cm$^{3}$ continuously traveling away from the Sun at speeds of the order of 500--1000 km/s   \cite{Biermann:1951ly}  \cite{Biermann:1957fk}. The suggested density was more or less consistent with the electron density inferred from the prevailing belief about the polarization of the diffuse zodiacal glow.} and Sidney Chapman's static solar atmosphere extending beyond the Earth.\footnote{Measurements of the temperature of the Earth's high atmosphere with rockets showed that the atmosphere became hotter at greater distances from the Earth. Chapman concluded that the heating must be caused by conduction of heat away from the Sun through the solar corona. The corona temperature of the order of million degrees, made it extend way out into interplanetary space, well beyond the orbit of Earth, with number densities of the order of $10^{2}$ ions and electrons/cm$^{3}$ at the orbit of Earth. But Chapman's\index{Chapman, Sidney} view was that of a {\it static} atmosphere, even if engulfing and affecting the Earth \cite{Chapman:1957bh}  \cite{Chapman:1959qf}.} 
  The conflict between the mutually exclusive ideas of Biermann and Chapman was actually solved by Eugene Parker who incorporated both concepts, and placed each in its proper perspective. He realized that the corona and the solar streams could not be separate entities, and suggested that the inward pressure of the interstellar medium was too weak to allow the solar atmosphere to be in hydrostatic equilibrium, but must expand continually into space filling the whole solar system, and generating a high-velocity radial flow with speeds of hundreds of kilometers per second, which he called ``solar wind''  to distinguish it from the older pictures of a static `atmosphere' or a bullet-like `corpuscular radiation'. He suggested that the weakening effect of gravity on the hydrodynamic flow would produce a transition from subsonic to supersonic flow.\footnote{Parker also expected that the expanding gas resulting from coronal temperatures of $\sim$ 2 million K would draw magnetic field lines of force out of the corona far into the solar system and, because of the solar rotation, the resulting interplanetary field would have a spiral pattern in the Sun's equatorial plane.}  
  
  Parker's paper   submitted  to the {\it Astrophysical Journal} in 1958  \cite{Parker:1958oq}, was initially rejected by two referees, and published in the November issue thanks to the then editor Subrahmanyan Chandrasekhar, as recalled later by Parker himself \cite{Parker:1997nx}.
  Many were skeptical. In criticizing Parker's solar-wind idea, Joseph Chamberlain's model, for example,  predicted a tenuous subsonic flow ---a ``solar breeze''--- for the expanding plasma \cite{Chamberlain:1960hc} \cite{Chamberlain:1961tg}.

 During the first meeting  of Rossi's Committee on Space Experiments held in September 1958, it became clear that  no experiment had been devised to enquire about the hydromagnetic conditions of space around our planet and  the Sun--Earth relation.  As Rossi remarked, <<it was hard to understand why this exploration should not have been included in the early program of NASA>> \cite[p. 131]{Rossi:1990aa}. After having reported the problem to the Space Science Board, Rossi felt thus motivated to initiate a study of interplanetary plasma  with his own group at MIT

Estimates of the bulk velocity varied from zero (stationary plasma) to about 1000 km/s (solar wind). However,  in case of a stationary plasma, protons would move at random, with expected kinetic energies of about 80 keV (in agreement with the temperature of 10$^{6}$ K, as observed in the visible corona). In the second case, the protons would move in a nearly parallel beam, with a kinetic energy of about 2.5 keV. The electrons, because of their higher thermal velocities, would in any case move more or less at random and therefore could not provide any information on the bulk motion of the plasma. Therefore, it was {\it necessary to study the proton rather than the electron flux in order to obtain information on the magnitude and the direction of the plasma flow}.

For the first time, physicists were tackling a totally new situation: experiments in space. In these early attempts it was  wise to sacrifice precision to simplicity of design, strength of construction and reliability of operation.  
For this reason, Rossi proposed to plan the detector  starting from the model of  a classical, well-tested instrument: the {\it Faraday cup}, a metal cup designed to catch charged particles in vacuum, which would measure the positive ion current, and by means of a repelling grid, give an indication of the ion energy. 

The most important quantities related to the interplanetary plasma were: spatial distribution of the plasma in the solar system and its variation with time, bulk velocity of the plasma, correlation with solar activity, correlation and association with magnetic fields, composition and  plasma ``temperature''. <<None of these quantities is known with any certainty,>> it was remarked in the first official presentation of the MIT plasma probe project signed by Rossi, Herbert S. Bridge (a ``pillar'' of the group in charge of the project), and by the young Frank Scherb.\footnote{LNS\&E {\it Progress Report}, May 31, 1959, p. 60.}

The uncertainties in the properties of the plasma, meant that a primary requirement of a plasma probe for the exploration of the interplanetary space was to be <<capable of giving information [\dots] on the magnitude and direction of the bulk velocity of the plasma over  as wide a range of values as possible.>> Once flying in space, the probe would have been  far from control of terrestrial laboratories,  so that they wanted it to be <<of solid design, simple in its construction and fully reliable in its operations>> \cite{Bridge:1960aa}. A collector plate  was to be mounted behind a hole in the skin of the space vehicle and grounded through a resistor. The hole was to be covered by a plane circular grid, G$_{4}$, kept at the potential of the vehicle's skin (to be called {\it zero potential}), while the collector plate was to be kept at a suitable negative potential. With the cup facing in the direction of the plasma stream, the plasma protons would flow through the grid to the collector, while the plasma electrons would be prevented from reaching the collector by its negative potential. The result would be a positive collector current, proportional to the proton flux.

\begin{figure}[ht]
\centering
\includegraphics [width=10cm]{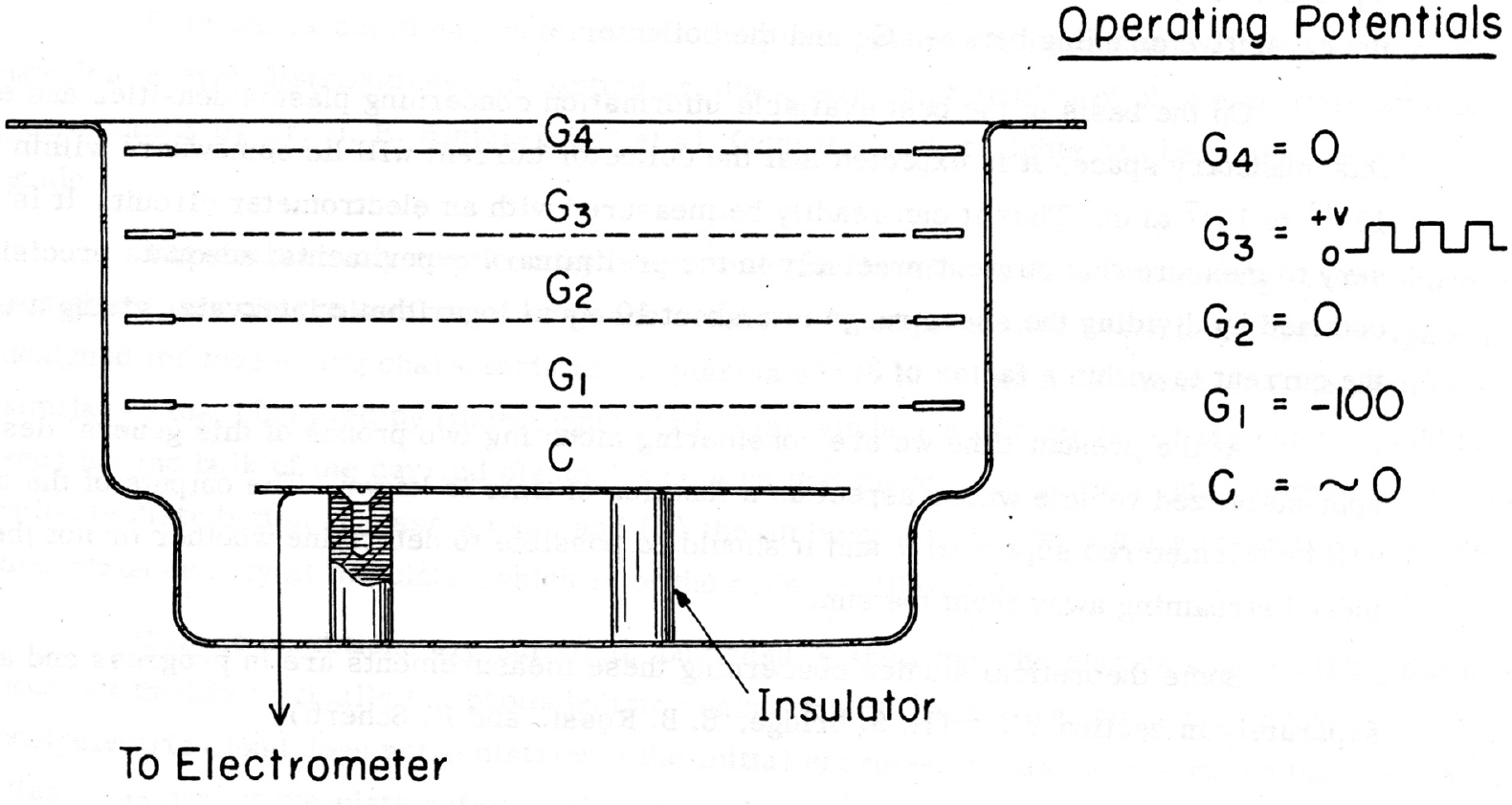}

\caption{Cross section of the MIT Plasma Probe, a multi-grid {\it Faraday Cup}.  The stream of plasma protons would flow through grid G$_{4}$ to the collector, kept at a suitable negative potential in order to prevent plasma electron from reaching it.  Grids G$_{1}$, G$_{2}$, and G$_{3}$ were added in order to fully remove all disturbances due to the direct and inverse photoelectric currents due to Sunlight impinging on the collector plate, and to  capacity coupling between the modulating grid and the collector. The collector plate as well as grids G$_{2}$, and G$_{4}$ were to be kept at zero potential with respect to the vehicle's skin [LNS\&E, {\it Annual Progress Report}, 31 May 1959,  p. 61].}\label{FaradayCup}
\end{figure}
\par\noindent

However, such a simple instrument was subject to a serious source of error: the strong solar radiation impinging upon the collector would eject electrons which, moving toward the grid, would produce in the collector circuit an electric current of the same sign as that produced by the protons moving in the opposite direction. A second grid, G$_{1}$, placed in front of the collector and kept at a negative potential relative to the collector would stop the  {\it direct} photoelectric current created by these electrons. While doing so, however, it would generate a {\it reverse} photoelectric current due to the solar radiation reflected toward it by the collector. They soon realized that it was not possible to entirely suppress the photoelectric disturbances by adding more grids, because each new grid would have been a new source of photoelectrons. As recalled by Rossi \cite[p. 134]{Rossi:1990aa}: <<Clearly, we had to contrive some entirely different method of making the response of our probe insensitive to the photoelectric effect. 
It was not an easy problem.>> They found a satisfactory solution  which consisted in {\it modulating} the photoelectric current using a third grid (the modulating grid G$_{3}$)  whose electric potential oscillated rapidly between zero and some adjustable positive voltage. The modulating grid, because of a positive square-wave imposed on it, would cut off the flow of protons with energies below a certain value which, for normal incidences, equalled the value of the modulating voltage. The modulation of the proton flow would produce in the detector circuit an alternating current of an intensity proportional to the modulated part of the proton flux. Photoelectric currents, not subject to modulation, would not affect the measurements. A further disturbance might arise from a capacity coupling between the modulating grid and the collector. Rossi himself solved this problem inserting another grid, G$_{2}$, directly behind the modulating grid.\footnote{A  thorough understanding of the performance of the instrument was a fundamental request. After building a model of the probe, the group undertook a comprehensive program of testing and calibration by placing the instrument in a vacuum system and exposing it to beams of protons of various energies, incident at various angles to the normal to the cup.}

While work on the plasma probe was in progress, a group of Soviet scientists led by Konstantin Gringauz had flown their detectors in several missions during 1959.\footnote{{\it Lunik 1} (launched January 2, 1959) was an attempt to reach the moon, but it missed by about 6000 km. {\it Lunik 2} (launched  September 12), was the first probe to hit the Moon. {\it Lunik 3}, launched on October 4 into a barycentric orbit, sent back the first pictures of the Moon's dark side.} 

In venturing into the interplanetary realm, {\it Lunik 2} and {\it Lunik 3} carried 4 ion traps each  with an inner grid held at $-$200 V to  repel photoelectrons produced  by incident radiation and an outer grid which was given different positive or negative potential with respect to the body of the satellite in order to obtain some information on the energy of the plasma particles  entering the cup.\footnote{The external grid was charged to a constant potential ranging from $-10$ to + 15 V on {\it Lunik 1} and {\it Lunik 2}, and from $-19$ to +25 V on {\it Lunik 3}. Actually, such a detector was not specifically designed as a plasma probe. The measured signal was indeed the sum of the fluxes of ions with energy above the outer grid potential, electrons above 200 eV, and photoelectrons produced at the electron suppression grid (<<The collector currents of the trielectrode traps may originate from charged particles of different nature>> \cite[p. 682]{Gringauz:1961zr}).}
 The Russian scientists were aware that  their design did not provide adequate protection against photoelectric interference, nor did it afford the possibility of significant measurements of the proton energies, the two important features that modulation  provided for the MIT probe.  However, even if it did not measure the speed or direction of the ions, it determined that there was indeed a flux of $\sim2\times10^{8}$cm$^{-2}$s$^{-1}$ of positive ions (presumably protons). From a comparison of the measurements made with different voltages on the outer grid, it appeared that the energy/charge $>$15eV/charge, corresponding to protons with  a speed of $>53$ km s$^{-1}$, indicating a particle density of 30 to 50 cm$^{-3}$. From these data they could thus conclude that they had observed for the first time   <<the corpuscular emission of the sun>>  in the interplanetary space outside the magnetic field of the earth \cite[p. 364]{Gringauz:1960ys}. 

Such a measurement favored Parker's theory, but left unanswered questions such as the extent to which the speed exceeded that limit, and the direction of the flow, even if it was claimed that  the signal fluctuated as the spacecraft spun around its axis, suggesting an ion flow was entering the instrument whenever it faced the Sun. Interesting results were also obtained with the Venus rocket launched in February 1961, which observed a flux of positive particles equal to about $10^{9}$ cm$^{-2}$s$^{-1}$ at a time when it was 1,890,000 km from the Earth \cite{Gringauz:1961zr}.

\begin{figure}[ht]
\centering
\includegraphics [width=14cm]{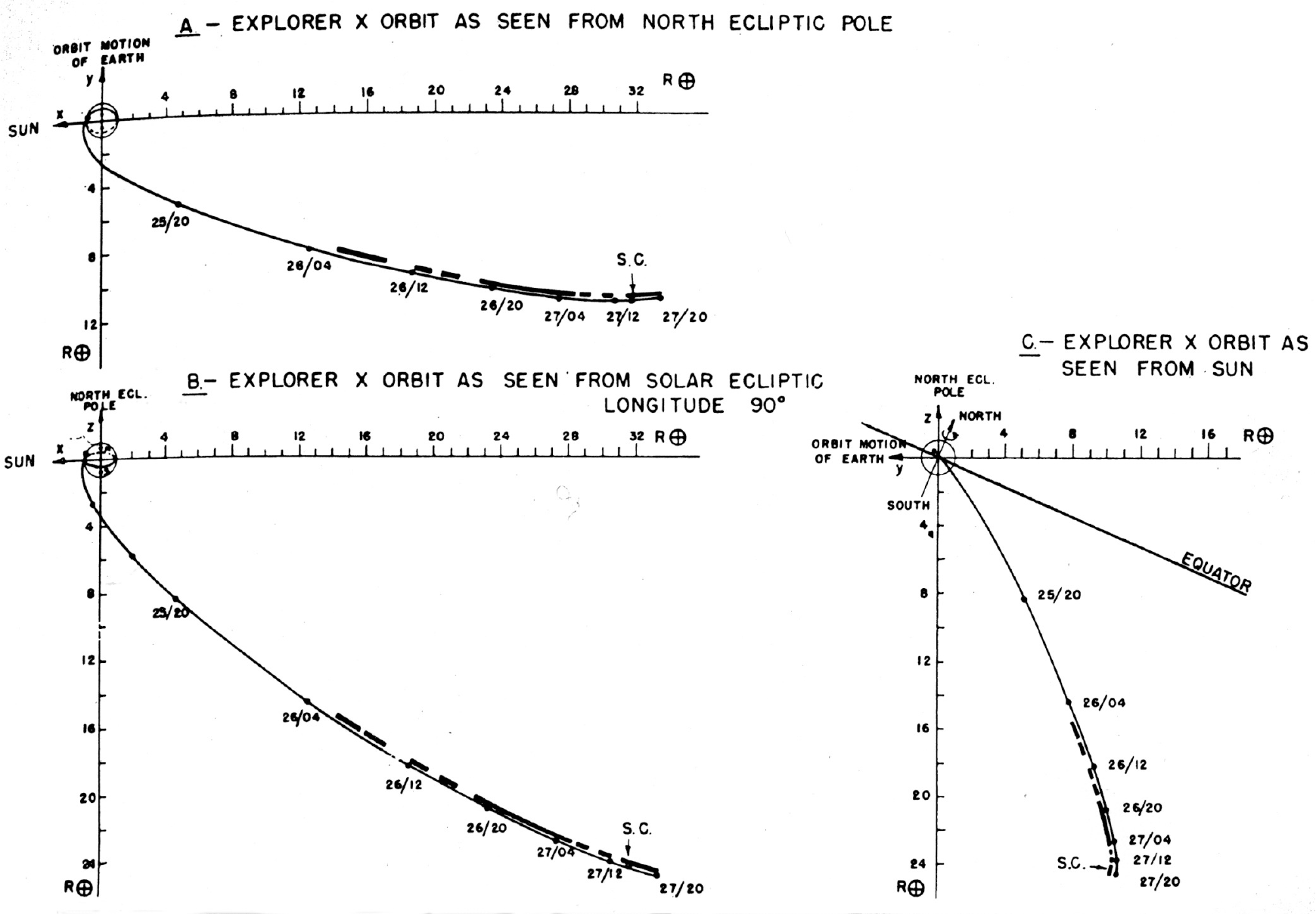}
\caption{Projections of {\it Explorer X} trajectory in the  $x-y$ plane (plane of the ecliptic), on the $x-z$ plane, and on the $y-z$ in solar ecliptic coordinates. Distances are in Earth radii. Numbers along trajectory are days and hours. Heavy black lines along projections indicate the sections of the trajectory where substantial fluxes of protons were detected  \cite{Bonetti:1963ab}.}\label{ExplorerXTrajectory}
\end{figure}
\par\noindent

\begin{figure}[ht]
\centering
\includegraphics [width=8cm]{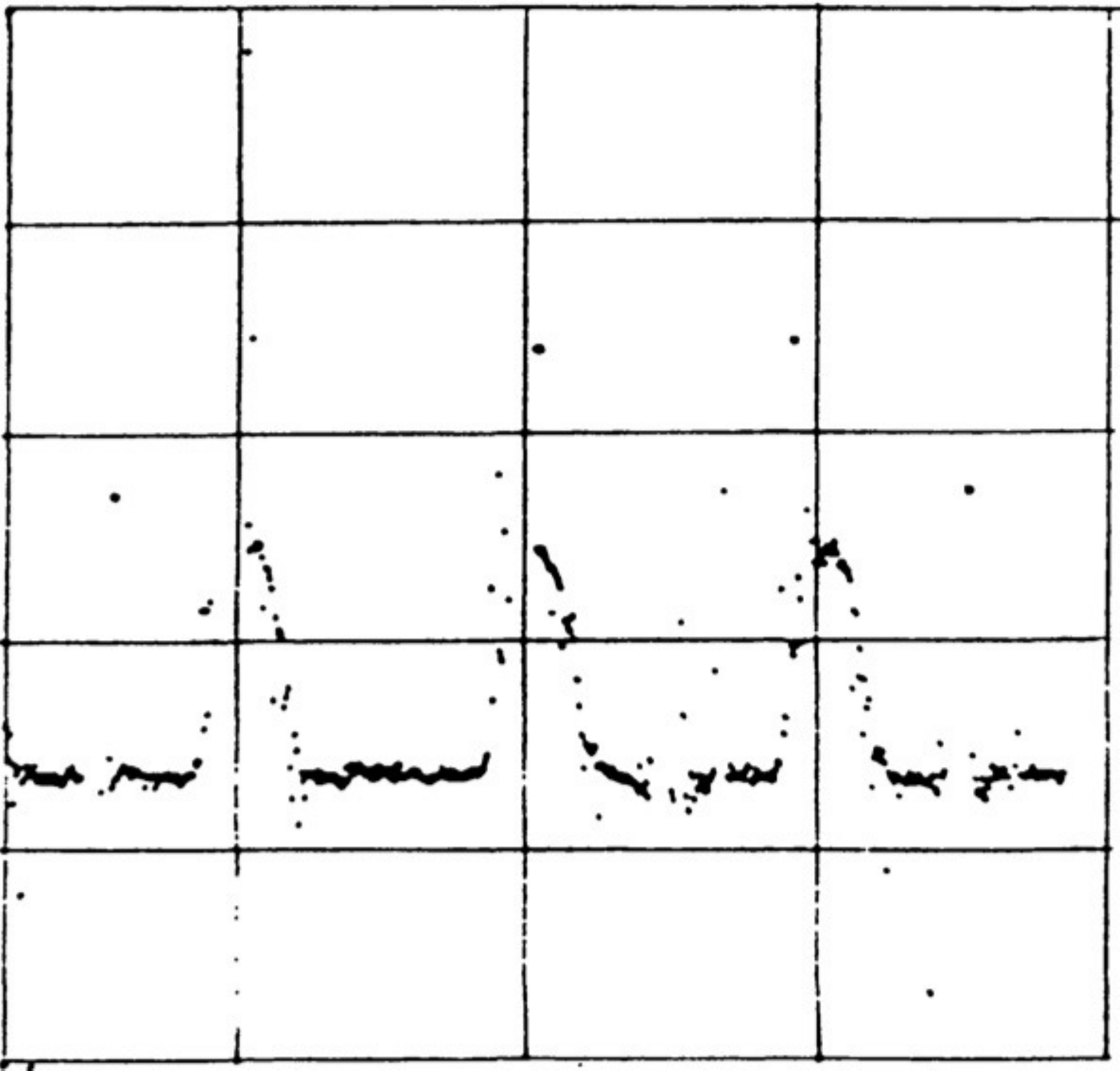}
\caption{Modulation of the plasma flux due to the rotation of the satellite {\it Explorer X}. Vertical lines show the azimuth of the Sun \cite{Bonetti:1963aa}.}\label{PlasmaFlux}
\end{figure}
\par\noindent

On March 25, 1961, a payload containing a rubidium vapor magnetometer, two fluxgate magnetometers, and the MIT plasma probe was launched from Cape Canaveral, Florida, and was injected into a highly elliptical orbit (perigee 300 km, apogee  240,000 km) surveying space on the night-side of the Earth. NASA named the successfully launched P-14 space probe \textit{Explorer X}. 
It was anticipated that if the plasma had a sufficiently high bulk velocity relative to the satellite, the detected proton flux would be modulated by the satellite rotation, reaching a maximum when the direction of the normal to the cup came closest to the plasma velocity vector. The sharpness of the maximum would provide a measure of the degree of collimation of the proton beam. Observations actually confirmed the prediction (see Fig. \ref{PlasmaFlux}).
The \textit{probe} measured  the ion flux in six different velocity intervals ranging from 120 up to 660 km/s (the maximum measurable velocity), with an average of about 300 km/s in the direction away from the Sun.
The number density  varied from about 6 to 20 protons/cm$^{3}$.

  The International Conference on Cosmic Rays and the Earth Storms held in Kyoto in September 1961 provided the opportunity of presenting some preliminary results  \cite[p. 555]{Bridge:1962fj}: <<Between 2.9 and 21.5 Earth radii no signal was observed [\dots] At about 21.5 Earth radii a flux of positive particles of about $4\times10^{8}$ cm$^{-2}$ sec$^{-1}$ at a mean energy of about 500 ev was observed for the first time. During the remainder of the flight, these particles were observed most of the time. However, there were large fluctuations in intensity [\dots]>>.  

When it was present, the magnetic field was weak and fluctuating; when the plasma was absent the field was strong and steady. The data were puzzling but consistent with the direction of motion of the plasma being radial away from the Sun. In any case, this direct observation was sustaining Parker's solar-wind theory. 

In the months following the Kyoto meeting, a detailed analysis of the difficult to interpret plasma measurements was carried out. The most significant finding was the existence of two sharply separate regions around the Earth. Although initially this had not been clear, the appearance and disappearance of a plasma together with the changes in the magnetic field were due to the fact the satellite, owing to its relatively low orbit, was moving almost along the boundary between  the relatively quiet region dominated by the Earth's magnetic field and interplanetary space, where solar wind is streaming around the magnetosphere. These results were thus materializing the existence  of a new frontier in the space surrounding the Earth.

The final results were presented at the Third International Space Science Symposium held in Washington, D. C., from April 30 to May 9, 1962   \cite{Bonetti:1963aa}  \cite{Bonetti:1963ab}. 
In a speech summarizing their conclusions, Rossi said \cite[p. 529]{Rossi:1963ly}:

\begin{quotation}
\small
Behind the Earth (i.e. downstream with respect to the plasma wind) there exists a region which is effectively shielded from the wind by the Earth's magnetic field. The boundaries of this region, which we may call `{\it geomagnetic cavity},' appear to be quite sharp. Beyond them, a plasma flow is observed, whose protons have a mean kinetic energy of about 400 eV, indicating for the plasma a bulk velocity of about 300 km s$^{-1}$. The proton flux fluctuates around a mean value of the order $3\times10^{8}$ particles cm$^{-2}$ s$^{-1}$. The direction of the plasma wind lies within a `window' of about 20$^{\circ}\times60^{\circ}$ aperture, which includes the Sun. An appreciable energy spread is observed, which may be explained by the assumption that the moving plasma has a `temperature' between 10$^{5}$ and 10$^{6}$ degrees Kelvin. {\it Explorer X} crossed the boundary of the geomagnetic cavity at a distance of about 22 Earth radii from the center of the Earth. However, on several occasions during the rest of the flight, the plasma current disappeared and then reappeared again. A tentative interpretation of this effect is that the satellite was flying close to the boundary of the geomagnetic cavity and that this boundary was not fixed in space but was moving back and forth, perhaps as a consequence of variations in the speed or in the density of the plasma wind.
\end{quotation}
\normalsize

  Actually, the reason why the velocity of the plasma wind observed by {\it Explorer X}  near the surface of the geomagnetic cavity was not very different from that of the wind unaffected by the presence of the Earth, was that {\it Explorer X}  was flying along the ``tail'' of the flapping magnetopause. It was clear that whatever interaction might occur in the vicinity of the Earth, it could not increase, but could only decrease the bulk velocity of the plasma. Therefore they concluded that the velocity of the undisturbed plasma must be at least as large as that which had been measured by {\it Explorer X}. Comparing the velocity of plasma flow with the phase velocity characteristic of plasma waves, they found that the wind was supersonic: 
  
  \begin{quotation}
\small
Under these circumstances, one should expect the formation of a `bow wave,' pointing toward the direction of the oncoming wind and enveloping the geomagnetic cavity.\footnote{They presented a schematic illustration of the geomagnetic cavity and the bow wave in \cite{Bonetti:1963ab}. The idea of a region from which clouds of ionized gas emitted from the Sun would be excluded by the action of the Earth's magnetic field  (the ``Chapman--Ferraro cavity'')  had been proposed by S. Chapman and C. A. Ferraro between 1930 and 1932 \cite{Chapman:1930lr} \cite{Chapman:1931aa}  \cite{Chapman:1932ac}.}
\end{quotation}
\normalsize

 The flight of {\it Explorer X} had thus established some of the basic facts about conditions prevailing in interplanetary space: the existence of a steady, albeit variable solar wind streaming past the Earth at supersonic speed, and the existence of a geomagnetic cavity, a region of space surrounding the Earth, which is shielded from the solar wind by the Earth's magnetic field. It prepared the stage for the complete vindication of Parker's theory of the solar wind.

In fact, what was really required, was a detailed study of the undisturbed solar wind from space on the day side of the Earth, or from interplanetary probes.  The {\it Mariner 2} mission launched on August 27 of that same year, carried a plasma probe built by Marcia Neugebauer and Conway Snyder of  Jet Propulsion Laboratory which could measure the energy spectrum of both electrons and protons. It observed for the first time the solar wind in free space  and definitely confirmed Parker's prediction of the supersonic expansion of the solar atmosphere \cite{Neugebauer:1962ys}.

New and important results concerning interplanetary plasmas and magnetic fields were obtained by means of vehicles reaching farther and farther into interplanetary space and carrying increasingly refined instruments. Rossi's direct involvement in plasma experimentation ended with Explorer X. For Herbert Bridge, instead, it was the start of a long and most fruitful research activity. The group that grew under Herbert Bridge's leadership, together with the theoretical group of Stan Olbert, placed MIT at the forefront in the investigation of interplanetary plasma.

\section{The X-Ray Window}

In parallel with his interest on plasmas in space, Rossi  had  also become convinced of the importance of exploring the X-ray window of the Universe. Already 
on July 1958, a few days after the first meeting of the Space Science Board, he wrote to Richard W. Porter (Chairman of the Technical Panel for the Earth Satellite Program, and member of the Space Science Board), expressing the opinion that <<Some exploratory work in the x-ray astronomy>> should be included in the satellite program for 1959--1960.\footnote{B. Rossi to Richard W. Porter,  11 July 1958, Rossi Papers, MIT Archives, Box 11, Folder ``Gamma Astronomy''.}
He suggested that <<the study of x-rays of the high-energy group be given first  priority in the early phase of satellite experiments. In this regard he mentioned <<a balloon-experiment program directed toward a preliminary study of x-rays from extraterrestrial sources,>> and added that they <<would be interested in undertaking a satellite experiment for the study of x-rays in the 10$^{8}$ ev region.>>\footnote{In the meantime, Kraushaar and Clark had continued to work on their $\gamma$-ray project, and on October 10, 1958, a proposal  for the support of a high-energy $\gamma$-ray satellite-borne experiment was submitted to the Space Science Board by the Cosmic-ray group of MIT.}

In 1959 the Sun was the only known celestial source of X-rays. It is a comparatively weak source, which can be detected easily from the Earth because of its proximity. For over ten years, solar X-rays had been  extensively and successfully studied by scientists at the Naval Research Laboratory  under the leadership of Herbert Friedman. Hard X-ray signals observed at lower altitudes softened as the rockets climbed through the atmosphere. A residual flux below 40 keV appeared to indicate a possible evidence of extrasolar X-rays. However, in the Spring of 1960, Friedman had to admit that attempts to observe X-ray extrasolar sources in the night sky had been unsuccessful up to that moment: <<It may be that the interstellar medium is too opaque to x rays to permit the observation of very distant sources>> \cite[p. 626]{Friedman:1960fk}.

Rossi, who was chairman of the Space Project Committee of the Space Science Board (``Analysis of advanced space research proposals and long-range planning''), had been invited to participate to the work of John Simpson's Committee for Physics of Fields and Particles in Space, whose Vice Chairman was James Van Allen.
In autumn 1959, in a document with experiments proposed for 1959--1960, Simpson's Committee recommended  that  <<A search should be made for continuous and burst X-radiation from the Sun, the Galaxy and from the Earth.>> However, they were aware of the great technical difficulty of revealing X-rays emitted from extrasolar sources, whose fluxes might be so low that it  could never be observed. 

 Evidence from ground-based radio astronomy had largely shown that a number of celestial objects were more abundantly emitting at radio wavelength, than in visible light.  It seemed reasonable to expect the radio spectra to continue through the visible and ultraviolet, all the way to X-rays. On October 24, 1958, the Interim Report of Simpson's Committee had specifically mentioned the interest in studies in the X-ray and $\gamma$-ray wavelengths. Lawrence H. Aller, a member of Leo Goldberg's Committee on Optical and Radio Astronomy, concluded a discussion on some aspects of ultraviolet satellite spectroscopy  remarking how interstellar absorption would be strong up to the region of 20 {\AA}, so that <<From distant stars and gaseous nebulae we can hope to observe only radiation in the x-ray region>> \cite[p. 329]{Aller:1959fk}. Leo Goldberg pointed out the importance of X-ray astronomy as a new research tool, but emphasized as well the great technical difficulty of revealing X-rays emitted from extrasolar sources, whose fluxes might be so low that it  could never be observed \cite{Goldberg:1959kx}.

 According to Martin Annis, Rossi's old friend and his former student at MIT, during a flight on the way home after a meeting of the Science Board, Rossi had reflected on the concrete possibility of revealing  X-rays of extrasolar origin: <<He was on the plane, sitting alone, and he thought that X-rays having the energy of some keV,  before being absorbed can go through a certain thickness of air  which can be compared to the mass absorption along galactic distances. The idea popped out in his mind. It was such a simple idea [\dots]>>\footnote{Interview of the author with M. Annis, September 30, 2006. Annis well remembers the circumstances, but does not remember the exact date of this conversation.\label{annis}} 

Soft X-rays  extend from the low-energy cut-off due to interstellar absorption to about 15 keV.
He was very excited  and  immediately talked with Annis about the matter after his  arrival: <<It just occurred to me that X-rays, specifically \emph{soft X-rays} will penetrate the entire Universe without absorption. We have never looked through the window of the soft X-ray region of the spectrum. I am convinced that exciting discoveries are to be made in this unknown world.>>\footnote{Interview of the author with M. Annis and F. Paolini, July 13, 2007. Annis had recently founded in Cambridge the {\it American Science \& Engineering} (AS\&E) company. At that time, Rossi was acting as a consultant for AS\&E as were, among others, George Clark and Stan Olbert. George Clark, too, has recalled how Rossi, coming back from a meeting of the Science Board, <<pointed out that there was a window beyond the hydrogen-helium absorption, and that X-rays in the energy range above a kilovolt should be transmitted freely through space, and that it might be worth looking for cosmic X radiation.>> Richard F. Hirsh, interview with G. W. Clark,  15 July 1976 (American Institute of Physics, Oral History Archives).\label{annispaoliniclark}}

Rossi's  basic ``exploratory philosophy'' was as usual very simple: <<[\ldots] since space is transparent to X-rays, and since one can think of a number of ways in which X-rays can be generated in space and by stars, one should go and look and see what you find.>>\footnote{Interview with G. W. Clark,  footnote \ref{annispaoliniclark}.}
  The very difficulty of detecting this type of radiation was clearly the kind of challenge that Rossi had always pursued during his whole scientific life \cite[pp. 146--147]{Rossi:1990aa}:
  \begin{quotation}
\small
  I was confident that, without resorting to any fundamentally new technology, it was possible to develop X-ray detectors substantially more sensitive than those previously used for solar observations. To be sure, the sensitivity of these detectors would still be a far cry from that deemed necessary for the detection of remote X-ray sources. But no one had yet explored the sky with X-ray detectors as sensitive as those that I hoped could be developed, and this, for me, was a sufficient reason for undertaking this exploration; my long experience as a cosmic-ray physicist had taught me that when one enters an unexplored territory there is always a chance that he may find something unpredicted.
 \end{quotation}
\normalsize

This awareness  convinced Rossi that it was worthwhile trying. As he had always done, he discussed the idea with his colleagues of the MIT Cosmic Ray-Group. But they were heavily committed to
other important research projects: the study of the interplanetary
plasma  and a search for $\gamma$-rays from celestial sources, which had
been initiated by William Kraushaar  and George Clark.\footnote{Notwithstanding Rossi's influence at MIT (<<A kind of Pope>>, according to Annis' definition) it cannot  be excluded that his project might have been rejected at the moment also because it sounded too hazardous to administrative spheres of MIT.} Still Rossi was <<not ready to forgo>> his plan \cite[p. 143]{Rossi:1990aa}, so he turned to Martin Annis and told him about exploring the existence of galactic sources of {\it soft X-rays}: <<Martin, nobody wants to listen to me or to work on this project, they don't let me do it\dots  What I want you to do Martin, is to develop optics to focus soft X-rays. And also I want you to develop detectors, large area detectors.>> Up to that time they had tiny detectors, remarked Annis, who added that he <<had never seen him such excited!>>\footnote{Interview with M. Annis, footnote \ref{annis}.}  

In talking with Annis, Rossi did not try to minimize the difficulties: <<He [Annis] fully
realized (as I did) that his fledgling company would take a considerable
financial risk by committing a sizable part of its facilities to a venture
of major proportion, without any apparent commercial applications.
Nonetheless, he accepted the challenge>> \cite[p. 144]{Rossi:1990aa}.

Rossi's intuition, that a new window of the electromagnetic spectrum could and should be opened thanks to space technology, thus inspired a program in X-ray astronomy at {\it American Science \& Engineering}, the company led by Martin Annis.\footnote{The company had worked on projects related to the Atomic Energy Commission program of atomic tests involving experiments for the measurements of X-rays and gamma rays from nuclear explosions, and had thus developed excellent capabilities for space missions. But Rossi felt that AS\&E should  redirect its work toward peaceful pursuits.} 

 \begin{figure}[ht]
\centering
\includegraphics [width=14 cm]{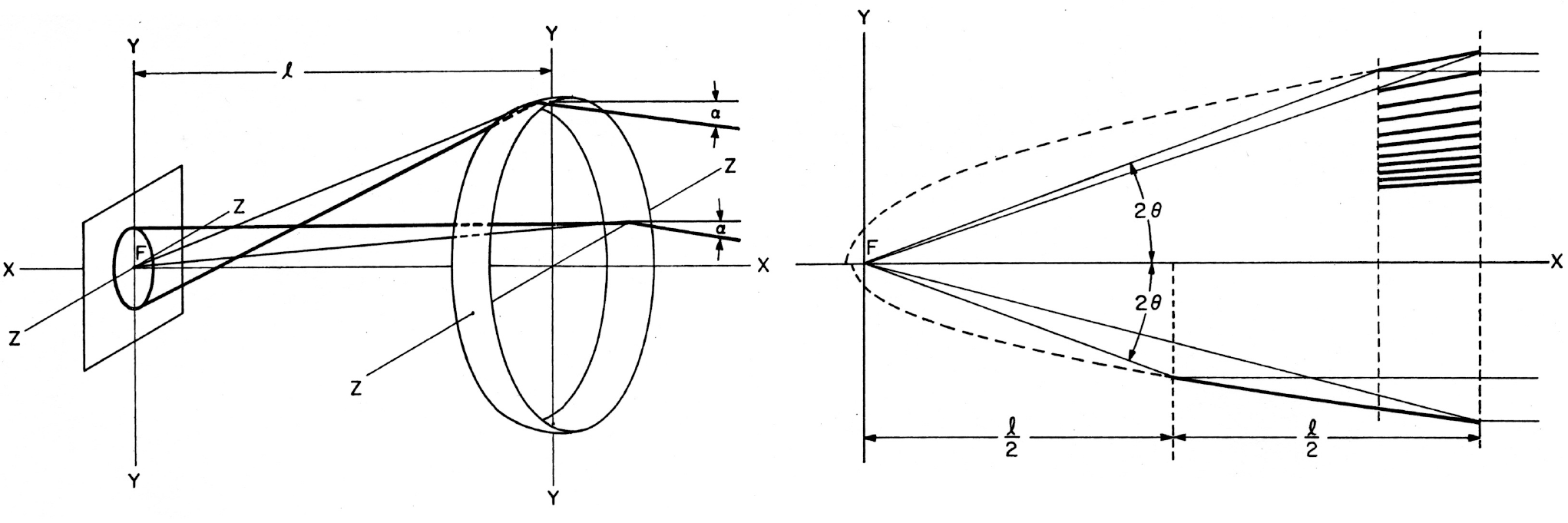}
\caption{Telescope for soft X-ray astronomy designed by  Giacconi and  Rossi. The angular resolution of the instrument is determined by the size of the detecting area at the focus of the paraboloid and by the perfection of the optical surface. Left: image formation by a small segment of a paraboloid; the incident rays are in the {\it xy} plane. Right:  Illustration of two possible solutions. The lower part of the figure is a section of a single parabolic surface; the upper part of the figure shows a section of a mirror composed of segments of confocal paraboloids   \cite{Rossi:1960fy}.\label{telescope}}
\end{figure}
\par\noindent

 The 28-year-old Riccardo Giacconi, a former student of Giuseppe Occhialini at the University of Milan, who had recently joined the Company in order to work to space projects, took charge of the program.  The challenge of developing an entirely new type of instrument was taken up, and a major effort aiming at the research of X-rays from celestial sources other than the Sun, was started in the fall of 1959. They were quite aware of the difficulties, as recalled by Giacconi (<<Considering the Sun to be a typical X-ray emitter, it became immediately obvious
that the expected flux at Earth from celestial sources would be below the threshold
of current detector systems by many orders of magnitude. This remained true even
when considering the possible emission from hotter stars or supernovae>> \cite[p. 4]{Giacconi:2005aa}), and by Rossi (<<We examined [with G. W. Clark and S. Olbert] the possibility that unusual X-ray objects, such as the Crab nebula or stars with high magnetic fields could be X-ray sources. And the results of this analysis were rather discouraging in the sense that, yes, there probably were X-ray sources, but very weak>>).\footnote{Richard F. Hirsh, interview with B. Rossi, 21 July 1976 (American Institute of Physics, Oral History Archives).\label{RossiInterview}}

A small group, initially formed by Rossi, Clark and  Giacconi  discussed the prospects for such a research and the necessity of achieving such greatly increased sensitivities in order to detect X-rays from such faint sources.\footnote{R. Giacconi, G. W. Clark, and B. Rossi, A Brief Review of Experimental and Theoretical Progress in X-Ray Astronomy, Technical Note of American Science \& Engineering, ASE-TN-49, January 15, 1960; R. Giacconi, G. W. Clark, Instrumentation for X-ray astronomy, ASE-TN-50, 15 January 1960; AS\&E proposal to NASA: Measurement of Soft X-Rays from a Rocket Platform Above 150 km, ASE-P-26, February 17, 1960.} 

Two parallel projects were developed: <<The first  key suggestion by Bruno was that we would have to develop an X-ray collection device to concentrate the flux.>>\footnote{Interview with M. Annis, footnote \ref{annis}.}   Giacconi studied the principles of X-ray optics developed by Hans Wolter in the late 1940s and suggested to use  total external reflection of X-rays under grazing incidence as the basic principle of the new instrument. Rossi suggested the refinement of nesting several mirror surfaces, thus increasing the effective collecting area and hence improve the detection rate \cite[p. 5]{Giacconi:2005aa}. The first line of attack was thus the development of an X-ray telescope  (Fig. \ref{telescope}) \cite{Rossi:1960fy},
  even if it was immediately clear that such a novel instrument would require a lengthy effort.\footnote{The paper by Rossi and Giacconi was received December 7, 1959 by the {\it Journal of Geophysical Research}. See also patent 3,143,651 filed on February 23, 1961 by Rossi and Giacconi, and patented 4 August 1964: \textit{X--Ray reflection collimator adapted to focus X-Radiation directly on a detector}. According to Annis, a prototype of the device was first built by Dr. Norman Harmon. M. Annis, personal communication, 9 August, 2008.}
  
  Again, all the people involved in this project were ``cosmic ray people'',  as Rossi himself recalled:\footnote{Interview with B. Rossi, footnote \ref{RossiInterview}.} 
  
\begin{quotation}
\small
  And the reason was partly practical that the detectors used in X-ray astronomy were counters not very dissimilar from the counters used in cosmic ray research, so these people were very familiar with that kind of instrumentation. The other reason is that astronomers are used to doing their observations with the instruments that other people build. They don't build their own telescopes. They're not accustomed to develop instruments, or at least not the major parts of the instruments which they are using in their experiments, like the cosmic ray people were. The first and most important reason is the one I mentioned before: that cosmic ray people were used to expecting the unexpected, while optical astronomers were not. 
\end{quotation}

  Meanwhile Annis, as president of AS\&E, had tried to get a contract for this research from  NASA. He was successful in having the Goddard Space Flight Center in  supporting the development of a prototype of the grazing incidence telescope, but the Agency was not willing <<to enter into the field because all the astronomers felt that the possibilities of success were exceedingly small.>>\footnote{Interview with B. Rossi, footnote \ref{RossiInterview}.}  At that time there were good reasons to suppose that the Moon should be a source of X-rays resulting from the impact of streams of energetic solar particles hitting its surface.
Therefore Annis turned to the   Lunar and Planetary Exploration Branch of the Air Force  Cambridge Laboratories, who <<looked favorably on a project aimed at detecting an X-ray emission by the Moon>> and agreed to finance the rocket instrumentation <<in the hope that it would provide some information on the chemical composition of the lunar surface.>>\footnote{Author interview with M. Annis and F. Paolini, footnote \ref{annispaoliniclark}.}

  The official purpose of the first experiment was thus to detect X-rays excited in the lunar surface by solar X-rays.   But the group, which now included also the young Herbert Gursky, had not renounced the main target: ``soft X-rays from galactic sources.''  
  However, in order to make in parallel a X-ray survey of the sky, hoping to observe the predicted fluxes from extrasolar sources, it was necessary to construct an instrument easy to build, using more or less conventional detectors such as GM-counters and proportional counters, but far more sensitive and sophisticated than those used previously, although necessarily less sensitive than a telescope.

 In this regard, a short term project was in fact underway, as recalled by Annis: <<A second key component of Bruno's   suggestions was that we would have to develop a large area detector with a very thin window for the detection of soft X-rays.  This was a very difficult task. Friedman used needle detectors for his prior experiments to detect X-rays from the Sun and Bruno was well aware that these detectors were not adequate for the project.>>\footnote{M. Annis, personal communication to the author, 9 August 2008.}
 
  \begin{figure}[ht]
\centering
\includegraphics [width=8cm]{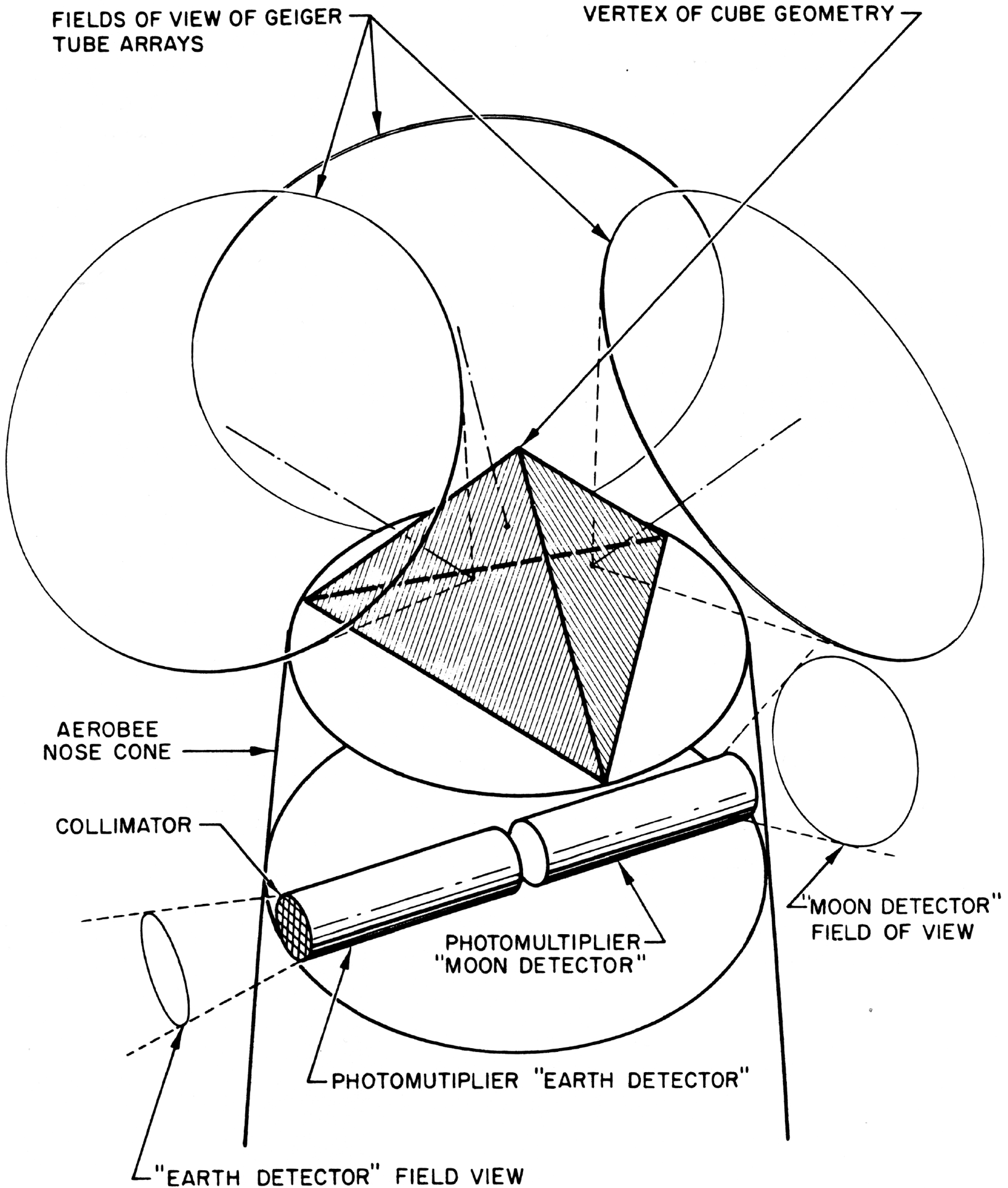}

\caption{Sketch of detector geometry allowing 2/3 of a hemisphere to be viewed at any one time. One of the three detector arrays comprising the X-ray detection system consists of a ``pancake'' GM-counter (or counters) filled with helium for the detection of X-rays of wavelengths 45 to 609 {\AA} alone. A second of the tube arrays is identical to the first, except for its filling, which is neon. This is for the detection of soft X-rays of wavelengths 2--15 {\AA} and 45--60 {\AA}. The third  detector array consists of a neon-or helium-filled ``pancake'' GM-counter having an aluminum window about 5 millimeters thick, which is opaque to the X-radiation of interest, but essentially transparent to cosmic radiation. The function of this detector array is the measurement of background counts which are not removed by the anti-coincidence shield. Each photomultiplier output is placed in anti-coincidence with the corresponding GM-counter array output. This arrangement is expected to eliminate about 90\% of the background counting rate arising from charged particles of the primary cosmic radiation flux or locally-produced showers. Original figure in ``Proposed Experiment for the Measurement of soft X-Rays from the Moon'' ASE-83-I (footnote \ref{ASE-83-I}).}\label{PancakeGeigerGeometry}
\end{figure}
\par\noindent
 
 \begin{figure}[ht]
\centering
\includegraphics [width=14cm]{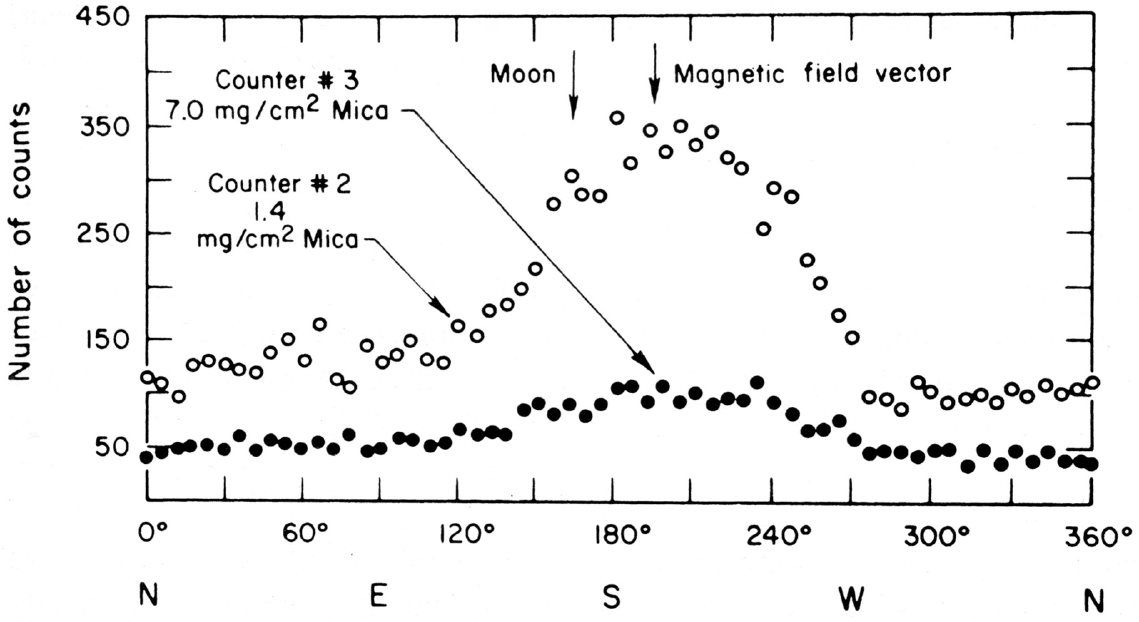}

\caption{The data provided by the Aerobee experiment of the AS\&E group flown on June 18, 1962. The number of counts versus azimuth angle, accumulated in 350 seconds in each 6$^{\circ}$ angular interval,  shows a clear peak for counter 2 at about 195$^{\circ}$, with a smaller peak for counter 3. The large peak shows that part of the recorded radiation is in form of a well collimated beam. The fact that the counting rate does not go to zero on either side of the peak shows that there is a diffuse background radiation  across the area of sky measured \cite{Giacconi:1962uq}.}\label{SCO-X1}
\end{figure}
\par\noindent

  The second line of attack had thus the aim of improving the performance of the thin-window gas counters already used  for solar X-ray astronomy. This work, which extended over a period of about two years, was carried out to a great extent by Frank R. Paolini, who worked at AS\&E as an expert of electronics. He   produced a detector about 100 times more sensitive than any of those flown previously,  conceiving a  system consisting of three arrays of ``pancake'' type GM-counters which together viewed a hemisphere of solid angle in order to increase the  chance of observing something during the flight (Fig. \ref{PancakeGeigerGeometry}).\footnote{The advantage of the pancake shape was the very large window-area-to-sensitive-volume ratio, a difficult achievement at that time. As stressed in a report of October 1960, the chosen detector geometry, in the presence of very strong X-ray sources other than the Moon, allowed <<the determination of the orientation of such sources to a much greater accuracy than might be supposed from consideration of the large field of view of the detectors alone. Because of the strong angular dependence of detector efficiencies, definite maxima in the Geiger tube counting rates may be observed as a strong source is swept. The time of occurrence of such maxima, together with aspect information, lead to the determination of the source orientation.>> ``Proposed Experiment for the Measurement of soft X-Rays from the Moon'', ASE-83-I, 25 October 1960, Rossi Papers, MIT Archives, Box 34.\label{ASE-83-I}}

In the words of George Clark, it was <<a very broad net, that they cast>>:\footnote{Interview with G. W. Clark,  footnote \ref{annispaoliniclark}.}
\begin{quotation}
\small
It was done in what I would say is the {\it typical Rossi exploratory style}, that always characterized his work that he has had any relation to, and that is, in the first place, {\it the imagination to go after something interesting}, and then, to do it {\it with a broad enough net} so you have a fair chance to catch it  [emphasis added]. 
\end{quotation}
\normalsize

Interest of some astronomers towards cosmic X-rays was growing.  The first conference on X-Ray Astronomy,  held on May 20, 1960, at the Smithsonian Astrophysical Observatory, Cambridge (Massachusetts), was attended by most of the research workers in this new field.\footnote{About half of the conference dealt with the theoretical astrophysical aspects of X-ray astronomy and the remaining half with the technical means for detecting extraterrestrial X rays. In talking about the instrumentation, Paul H. Kirkpatrick discussed an  X-ray telescope based on the grazing incidence principle, similar to the instrument designed by Giacconi and Rossi.} As recalled in the introductory talk by Robert Davis, <<If we look for radiation from, say Alpha Centauri, which is very similar to the sun, we would find the received fluxes to be ten orders of magnitude smaller, even without considering interstellar absorption.>> However, it was clear that one could have good reason to believe that <<other types of objects are X-ray emitters, for example, the Crab Nebula, where strong emission occurs in both the visible light and radio regions [\dots] It's fairly safe bet to presume that radiation extends into the X-ray region>> \cite[p. 10]{Berman:1960uq}. On the other hand, Davis remarked how 
Aller's assumption \cite{Aller:1959fk} that <<interstellar space begins to be transparent at 20 {\AA} is perhaps a bit too optimistic; perhaps it occurs at a somewhat lower wavelength [\dots]>> In any case, the general discussion could not but examine more or less intense fluxes of X-rays produced by what at the time were considered standard processes taking place in stars and interstellar regions. The possibility that other astrophysical sources might emit copious fluxes of X-rays by production mechanisms which could not be foreseen was put forward by Stan Olbert, of the MIT Cosmic Ray Group, who had been requested by Rossi to do some theoretical work on X-ray production by various stellar and interstellar processes: <<I believe that when we begin detecting and studying X rays from various astronomical sources, we will be in for a number of surprises. We may discover that the fluxes may be higher than we anticipate, coming from a number of different processes [\dots]>>  \cite[p. 45--46]{Berman:1960uq}.

And indeed Rossi's restlessness, which had always made him move from one problem to another always hoping to detect some yet unknown aspect of Nature, was to be once more satisfied. After two failures a new rocket launching of the AS\&E payload took place at the White Sands Missile Range, New Mexico, on June 18, 1962, a day after full moon. The rocket payload carried an optical aspect sensor, a magnetometer and three of Frank Paolini's counters provided with windows of varying thickness so that the energy of the radiation could be detected. They were mounted symmetrically on the side of the rocket which rose to a maximum altitude of 225 km, while rotating at the rate of 2 revolutions per second around its spin axis. At each revolution the counters scanned a portion of the sky 100$^{\circ}$ wide. Throughout the flight the axis was pointing approximately in the vertical direction. At an altitude of 80 kilometers, the rocket having reached beyond the dense layers of the atmosphere, an increase in the pulse frequency of the counters signaled the appearance of a rather strong flux of radiation. The signal was strongly modulated by the rotation of the rocket, evidence that the X-rays did not come in equal numbers from all directions. The rocket remained above 80 kilometers for a total of 350 seconds. Of the three counters, two worked reliably.

The  dots in Fig. \ref{SCO-X1} show the total number of counts recorded during he entire flight. The higher curve is outlined by points representing data accumulated by  counter  2 as a function of the azimuthal angle of the direction perpendicular to its plane. The lower curve was constructed with data obtained by the  counter 3 provided with a thicker window (about 1 millimeter), and therefore less sensitive to X-rays.

The analysis of the records was not easy, as Rossi recalled later, they spent weeks and weeks  to make sure that there was nothing which might have misled them in the interpretation of their observation, and to convince themselves that the X-ray source wasn't the moon actually: <<We were all skeptical at the beginning, and a lot of work was done to try to evaluate all possible sources of error.>>\footnote{Interview with B. Rossi, footnote \ref{RossiInterview}.} 

The investment of more than two years of hard work  had returned only about 5 minutes of data, but the reward made all the effort worthwhile. At the end of  summer they became sure that the data could <<best be explained by identifying the bulk of the radiation as soft X-rays from sources outside the solar system>> \cite{Giacconi:1962uq}.\footnote{As far as the position of the source was concerned, Paolini recalled that he was the only one at As\&E at the time who knew how to work with celestial coordinates. According to his analysis he concluded that the source was in the constellation of Scorpius. He was thus convinced that the source was an extrasolar one. But unfortunately <<conservatism prevailed> and they never mentioned Scorpius in their early publications. Frank Paolini to Herbert Bridge, 20 December 1976. I am very grateful to F. Paolini for sending me a copy of this letter in May 2007. Many years later, in writing his biography, Bruno Rossi acknowledged Paolini's analysis: <<I regret that I did not encourage him to publish this result>> \cite[p. 151]{Rossi:1990aa}.}

The experiment had also proved that the Universe contains background radiation of X-ray light.

The discovery of the first extra-solar X-ray source, was presented August 1962 in occasion of a congress at Stanford University and was  welcomed with enthusiasm, though mixed with considerable skepticism. It appeared incredible that a very simple instrument could have revealed such an intense signal as to appear incompatible with the available astronomic data {\it regarding the celestial bodies then known}. It is worth noting that the article presenting those results was published by the {\it Physical Review} only after Rossi assumed personal responsibility  for the assertions thereby contained \cite[p. 151]{Rossi:1990aa}. Thirty years had elapsed since the young Bruno Rossi had to request the authority of Werner Heisenberg to have the plausibility of his early scientific results validated\dots\footnote{See footnote \ref{rossiheisenberg}.}

The experiment was repeated in exactly the same conditions in October 1962, when the presumed source was below the horizon, and again in June 1963, when the expected maximum reappeared in exactly the same position of one year earlier. In April 1963 the source was  located  by C. Stuart Bowyer and Herbert Friedman  in the constellation of Scorpio \cite{Bowyer:1964fk}, hence the name of Scorpius X-1 by which it became known.\footnote{They used a new detector with a window area of 65 cm$^{2}$ which was ten times as sensitive as the detector of the AS\&E group. Moreover, it was able to measure much more accurately the positions of possible sources.} They also detected X rays from  the Crab Nebula \cite{Bowyer:1964ab}

 Scorpius X-1 was  some 9,000 light years away and apart from the Sun, it revealed the most powerful X-ray source in Earth's skies, an object emitting a thousand times more X rays than the Sun at all wavelengths and a thousand times more energy in X rays than in visible light. An entire new astronomy was opening,  an essential source of information about new classes of stellar objects,  which since then has thrown light in previously unknown processes going on in the Universe.

AS\&E\index{American Science \& Engineering} took the lead for many years in developing X-ray astronomy, a  field which had a giant development also thanks to Riccardo Giacconi  who was awarded the 2002 Nobel Prize for Physics <<{\it for pioneering contributions to astrophysics, which have led to the discovery of cosmic X-ray sources}.>>\footnote{Giacconi shared one-half of the Prize  with Raymond Davis Jr. and Masatoshi Koshiba  who got it <<for pioneering contributions to astrophysics, in particular for the detection of cosmic neutrinos.>>}

His mentor Bruno Rossi, who was also deserving of that Nobel Prize,   was no longer alive.

\section{Conclusion}

Rossi's intuition on the importance of exploiting  new technological windows to look at the universe with new eyes is  a fundamental key to understand the wide underlying theme which shaped the natural evolution of his scientific identity  from  a ``cosmic-ray physicist'' to a ``cosmic-ray astronomer''. His scientific path is strongly interlaced with cosmic rays as a key to understand natural phenomena first at a microscopic level, and later at a cosmic scale. The new field of astroparticle physics, which emerged at the intersection of particle physics, astronomy and cosmology, is thus providing the proper perspective to look at  Rossi's scientific path. 

A profound unity  characterizes the work of Bruno Rossi, guiding his  research line up to the culminating moment of his scientific career at the beginning of 1960s when he and his group at MIT demonstrated the existence of the solar wind and when he promoted the search for extra-solar sources of X rays, an idea which eventually led to the breakthrough experiment which marked the inception of cosmic X-ray astronomy,   today one the main instruments to investigate astrophysical processes and the nature of the celestial bodies generating them. These achievements became  the final acts of his coherent research project, begun on the hills of Arcetri thirty years before: 
\begin{quotation}
\small
{\it Whenever technical progress opened a new window into the surrounding world, I felt the urge to look through this window, hoping to see something unexpected.} 
\end{quotation}
\normalsize

This lifelong urge and the consciousness {\it that the riches of nature far exceed the imagination of man},  make Bruno Rossi  one of the last  ``natural philosophers''  of the 20th century, in the grand tradition of Galileo Galilei.

\section*{Acknowledgments}
I am very grateful to an anonymous reviewer for some useful comments.

%\nolinenumbers   

\bibliographystyle{plainnat}

\end{document}